\pdfoutput=1
\documentclass[11pt,a4paper]{article}

\usepackage{MySty}
\usepackage{amsmath,amsfonts,amssymb,graphics}
\usepackage[normalem]{ulem}
\usepackage{graphicx}
\usepackage{mathrsfs}
\usepackage{bbm}
\usepackage{pifont}
\usepackage{braket}
\usepackage{footmisc}
\usepackage[dvipsnames]{xcolor}
\usepackage{mathtools}
\usepackage{xspace}
\usepackage{booktabs}
\usepackage{siunitx}
\usepackage{slashed}
\usepackage{cprotect}
\usepackage{tikz}
\usetikzlibrary{positioning,shapes,arrows,calc}
\usetikzlibrary{decorations.pathmorphing}
\usetikzlibrary{decorations.markings}
\usepackage{standalone}
\usepackage{csquotes}
\usepackage{multirow}

\graphicspath{{figs/}}

\title{ \bf{Kick it like DESI:}\\
\texttt{PNGB quintessence with a dynamically generated initial velocity}}
\author{Maximilian Berbig}
\affiliation{Departament de Física Teòrica, Universitat de València, 46100 Burjassot, Spain,\\
Instituto de Física Corpuscular (CSIC-Universitat de València), Parc Científic UV,\\
C/Catedrático José Beltrán, 2, E-46980 Paterna, Spain}
\emailAdd{berbig@ific.uv.es}

\abstract{Motivated by the hint for time-dependent dynamical dark energy from an analysis of the DESI Baryon Accoustic Oscillation (BAO) data together with information from the Cosmic Microwave Background (CMB) and Supernovae (SN), we relax the assumption of a vanishing initial velocity for a quintessence field. In particular we focus on pseudo-Nambu-Goldstone-Boson (PNGB) quintessence in the form of an axion like particle, that can arise as the phase of a complex scalar and could possess  derivative couplings to fermions or topological couplings to abelian gauge fields, without upsetting the necessary flatness of its potential.  We discuss mechanisms from the aforementioned interactions for sourcing an initial axion field velocity $\dot{\theta_i}$ at redshifts $3\leq z\leq 10$, that will \enquote{kick} it into motion.  Driven by this initial velocity the axion will first roll up in its potential, similar to \enquote{freezing} dark energy. After it has reached the pinnacle of its trajectory, it will start to roll down, and behave as \enquote{thawing} quintessence. As a proof of concept we undertake a combined fit to BAO, SN and CMB data at the background level. We find that a scenario with $\dot{\theta_i}=\mathcal{O}(1) \; m_a$, where $m_a$ is the axion mass, is slightly preferred over both $\Lambda$CDM and the conventional \enquote{thawing} quintessence with $\dot{\theta_i}=0$. The best fit points for this case exhibit transplanckian decay constants and very flat potentials, which both are in tension with conjectures from string theory.}

\begin{document}
\maketitle

\section{Motivation}
\begin{figure}[t]
    \centering
    \includegraphics[width=0.75\textwidth]{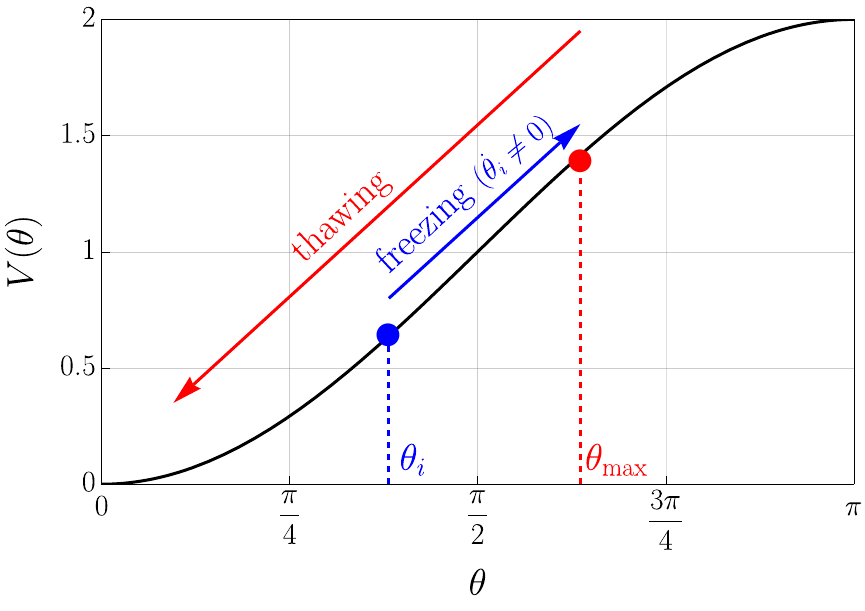}
    \caption{Schematic illustration of the evolution of the axion quintessence field with a non-vanishing initial velocity $\dot{\theta_i}$ in its potential $V(\theta)$. The field starts out at $\theta_i$ and gets pushed up its potential by its initial velocity (blue arrow), until it reaches $\theta_\text{max}$ (see eq.~\eqref{eq:Max}), which is similar to \enquote{freezing} quintessence. Then, once it has reached $\theta_\text{max}$, it starts to roll down towards its minimum (red arrow), which corresponds to \enquote{thawing} quintessence. The chosen values for $\theta_i$ and $\theta_\text{max}$ are for visual reference only. Here we chose $\dot{\theta_i}>0$, which is preferred by our  fit to cosmological data (see section \ref{sec:BOTH}).}
    \label{fig:diagram}
\end{figure}

What started out more than one hundred years ago as the self-proclaimed blunder of one of mankind's greatest thinkers, whose insights profoundly revolutionized our understanding of nature, is still one of the greatest mysteries in modern day cosmology: the cosmological constant (see Refs.~\cite{Weinberg:1988cp,Carroll:1991mt,Carroll:2000fy} for reviews on the subject). Since the late 90s the cosmological constant is back on the menu \cite{Krauss:1995yb}, because of  observational evidence for the accelerated expansion of space from measurements of the brightness-redshift relation of Type I Supernovae (SN) \cite{SupernovaCosmologyProject:1997zqe,SupernovaSearchTeam:1998fmf,SupernovaSearchTeam:1998bnz,SupernovaCosmologyProject:1998vns}.

More recently large scale structure data from Baryon Accoustic Oscillations (BAO) measured by the \verb|DESI| collaboration \cite{DESI:2024hhd} hints at the possibility that the expansion is not driven by an actual constant but rather by a time dependent dark energy. Such a behavior can be most easily accommodated by the evolution of a scalar field background, which is most commonly referred to as quintessence \cite{Fujii:1982ms,PhysRevD.35.2339,Wetterich:1987fm,Ratra:1987rm} (see Ref.~\cite{Copeland:2006wr} for a review).
Of course this presumes a mechanism for the absence of the bare vacuum energy.
Such mechanisms include, but are not limited to, the cancellation or relaxation of the cosmological constant \cite{Abbott:1984qf,Barr:1986ya,Peccei:1987mm,Barr:1988zk,Rubakov:1999aq,Hebecker:2000au,Barr:2006mp,Alberte:2016izw} (which has to evade the \enquote{no-go} theorem put forward in  Ref.~\cite{Weinberg:1988cp}), large extra dimensions \cite{Arkani-Hamed:2000hpr}, formal arguments from string theory  \cite{Obied:2018sgi,Garg:2018reu,Ooguri:2018wrx} or the S-matrix formulation of quantum gravity  \cite{Dvali:2013eja,Dvali:2014gua,Dvali:2017eba,Dvali:2020etd}, that both do not allow  a de Sitter phase.

One well studied candidate  for the quintessence field are Pseudo-Nambu-Goldstone-Bosons (PNGB) \cite{Frieman:1995pm,Choi:1999xn,Nomura:2000yk,Kim:2002tq,Hall:2005xb,Barbieri:2005gj} defined in terms of   compact dimensionless angular fields $\theta \in [-\pi,\pi]$, which are ubiquitous in string theory \cite{Svrcek:2006yi,Arvanitaki:2009fg}.
The  inherent shift symmetry of these  fields, which we will generically refer to as axions, prevents additional quantum corrections to their symmetry-breaking potential of the form $m_a^2 f_a^2(1-\cos{(\theta))}$, where the UV scale $f_a$ is known as the axion decay constant. This potential is required to be incredibly flat since the axion mass $m_a$ has to be comparable to the Hubble rate today of  $H_0=\mathcal{O}(10^{-33}\;\text{eV})$. Unlike non-compact scalars, \cite{Carroll:1998zi} PNGBs are essentially free from constraints due to fifth-force searches \cite{Fischbach:1985tk} (consider also Ref.~\cite{Fischbach:1999bc} for an overview), as a spin-independent long range force can only be mediated by exchanging two PNGBs and the resulting potential between test bodies scales with distance $r$  as $V(r)\sim 1/r^5$ \cite{Grifols:1994zz,Ferrer:1998ju}\footnote{An exception occurs, when the PNGB obtains its mass from gluons, which leads to $V(r)\sim 1/r^3$ \cite{Bauer:2023czj}.}. Thus axions from e.g. the phase of a complex scalar field, with their derivative couplings to fermions or topological couplings to abelian gauge fields, are a viable possibility for quintessence. 

Unlike attractor models \cite{Zlatev:1998tr,Steinhardt:1999nw} PNGB quintessence  is typically dependent on the initial conditions\footnote{Models with potentials of the form $(1-\cos{(\theta))}^n$ with $n<0$  can have   attractor solutions \cite{Hossain:2023lxs,Boiza:2024azh}.} for the  position $\theta_i$ and the initial velocity $\dot{\theta_i}$, which is usually presumed to be vanishing $\dot{\theta_i}=0$.
However as the data seems to suggest a non-vanishing kinetic energy of the dark energy today, one could ask what happens in scenarios, where the axion starts out its evolution with an initial kinetic energy.
In this work we exploit the aforementioned couplings to generate a small velocity for the axion field, that is injected at redshifts $3\leq z \leq 10$ and \enquote{kicks} it into motion. Usually a quintessence field rolls down its potential, which is called \enquote{thawing} quintessence \cite{Scherrer:2007pu}. In our proposal on the other hand, it gets first pushed up in its potential due to the velocity, similar to scenarios known as \enquote{scaling} or \enquote{freezing} quintessence \cite{Ferreira:1997hj,Copeland:1997et,Liddle:1998xm,Dutta:2011ik}, before it stops and begins it descend, while acting as \enquote{thawing} quintessence. This chain of events was sketched in figure \ref{fig:diagram}.

We fit the Hubble rate predicted by this model to data from \verb|DESI| BAO  \cite{DESI:2024hhd}, cosmic microwave background (CMB) observables from  \verb|Planck|  \cite{Planck:2018vyg,Planck:2019nip} and the Atacama telescope \verb|ACT| \cite{ACT:2023dou,ACT:2023kun} as well as the SN dataset \verb|Pantheon+|  \cite{Scolnic:2021amr} following the procedure outlined in Ref.~\cite{Wolf:2024eph}. From this analysis we find that \enquote{kicked} axion quintessence offers a slightly better fit to the data than both the cosmological concordance model $\Lambda$CDM or thawing quintessence. The resulting transplanckian values of $f_a$ and the flatness of the potential are unfortunately in tension with conjectures motivated by quantum gravity and string theory.

This manuscript is structured as follows:
Section \ref{sec:data} gives an overview over the current hints for dynamical dark energy in the data. In section \ref{sec:PNGB} we first provide a pedagogical review of PNGB quintessence models with $\dot{\theta_i}=0 $, and argue that the slow roll approximation can be relaxed.
Readers already familiar with the subject are advised to proceed directly to section \ref{sec:Kick}, which introduces the effect of a non-vanishing $\dot{\theta_i}$.  Possible models that can source the required a velocity are summarized in section \ref{sec:models}, as well as in the appendices \ref{sec:ooE}-\ref{sec:HelGauge}. The numerical fit to data is explored in sections \ref{sec:num}-\ref{sec:discuss}.
We discuss constraints from fragmentation of the homogeneous PNGB condensate and the implications for conjectures motivated by string theory in section \ref{sec:constr}, before concluding in  \ref{sec:conc}.

\section{DESI results and CPL-Parameterization}\label{sec:data}
Time dependent dark energy is typically fit by using the
Cheavllier-Polarsky-Linder (CPL) \cite{Chevallier:2000qy,Linder:2002et} parameterization, which reads  
\begin{align}\label{eq:EOSCPL}
    \omega_\text{CPL}(z) = \omega_0 +\frac{z}{1+z} \omega_a,
\end{align}
and implies the following Hubble rate for a spatially flat universe
\begin{align}\label{eq:HubCPL}
    H(z) = H_0 \sqrt{\Omega_m (1+z)^3 + \Omega_r (1+z)^4 + \left(1-\Omega_m - \Omega_r\right) e^{- \frac{ 3 \omega_a z}{1+z}}\left(1+z\right)^{3(1+\omega_0+\omega_a)}},
\end{align}
in terms of the matter and radiation density parameters $\Omega_m$, $\Omega_r$. 
Combining \verb|DESI| BAO data \cite{DESI:2024hhd}, CMB observables from  \verb|Planck|  \cite{Planck:2018vyg,Planck:2019nip} and the Atacama telescope \verb|ACT| \cite{ACT:2023dou,ACT:2023kun} as well as the SN dataset \verb|Pantheon+|  \cite{Scolnic:2021amr} implies for a spatially flat universe \cite{DESI:2024mwx}
\begin{align}\label{eq:DESIlimits}
    \omega_0 = -0.827 \pm 0.063 ,\quad \omega_a =-0.75^{+0.29}_{-0.25}
\end{align}
and combining this with the full shape power spectrum of galaxy, quasar and Lyman-$\alpha$ tracers \cite{DESI:2024aax} results in \cite{DESI:2024hhd}
\begin{align}
    \omega_0 = -0.858  \pm 0.061, \quad \omega_a = -0.68^{+0.27}_{-0.23}.
\end{align}
The impact of the choice of the SN data sets was discussed in Ref.~\cite{Efstathiou:2024xcq} (see the discussion in section \ref{sec:SN}, which explains why we focus on \verb|Pantheon+| data), the choice of BAO data in \cite{Chan-GyungPark:2024spk}   and the choice of CMB data sets in \cite{Giare:2024ocw}. Effects due to varying the CMB lensing consistency parameter were assessed in \cite{Chan-GyungPark:2024brx,Chan-GyungPark:2025cri}.
A fit of  the CPL parameterization is in general very sensitive to both  the choice of priors \cite{Cortes:2024lgw} and the choice of the dataset 
\cite{Wolf:2023uno,Wolf:2024eph} via the earliest included redshift. 

At the level of the CPL parameters one can already see, that the data seems to prefer an equation of state that crosses below $-1$ at early times, since $\omega_0 + \omega_a <-1$. This regime is known as the \enquote{phantom} regime \cite{Caldwell:1999ew,Caldwell:2003vq}, and it can most economically be realized via a scalar field with a wrong sign kinetic term. These constructions typically violate the Null Energy Condition of General Relativity \cite{Qiu:2007fd} and suffer from various pathologies such as vacuum decay \cite{Cline:2003gs}; a review can be found in Ref.~\cite{Ludwick:2017tox}. Even so it is known, that well behaved, canonical  quintessence models, which do not exhibit phantom behavior ($\omega(z)\geq -1, \; \forall z$), can be mapped to regions in the $\omega_a$ versus $\omega_0$ parameter space featuring a phantom crossing \cite{Linder:2006xb,Wolf:2023uno,Shlivko:2024llw}. The authors of Ref.~\cite{Gialamas:2024lyw} argue that the apparent preference for phantom behavior is merely an artifact from extrapolating the CPL parameterization to large redshifts.

\section{PNGB quintessence}\label{sec:PNGB}
We consider a compact pseduoscalar $a$ with a canonical kinetic term, that can be the PNGB of a global $\text{U}_\text{X}(1)$ symmetry with a potential of the form 
\begin{align}\label{eq:Va}
    V(a)= m_a^2 f_a^2 \left(1-\cos{\left(\frac{a}{f_a}\right)}\right).
\end{align}
Here $f_a$ is the axion decay constant given by e.g. the order parameter that spontaneously breaks the $\text{U}_\text{X}(1)$ and the potential can be induced by an explicit breaking, either via gravitational effects or a confining gauge symmetry, that has a mixed anomaly with $\text{U}_\text{X}(1)$. More comments on the potential and its origin will be provided in section \ref{sec:QG}.
We chose the minimum of the potential $a=0$ in order to ensure the absence of a cosmological constant. 
In the following we call the PNGB an axion and further set
\begin{align}
    \theta \equiv \frac{a}{f_a}.
\end{align}
In order to act as the quintessence field driving the exponential expansion of the universe, the PNGB equation of state in terms of the energy (pressure) density  $\rho_\theta\;(P_\theta)$
\begin{align}
    \omega = \frac{P_\theta}{\rho_\theta}= \frac{\frac{\dot{\theta}^2 f_a^2}{2}-V(\theta)}{\frac{\dot{\theta}^2 f_a^2}{2}+V(\theta)}
\end{align}
has to be smaller than $-1/3$. A cosmological constant would lead to $\omega = -1$, but a dynamical scalar field has a non-vanishing kinetic energy, which will lead to a time dependent equation of state.
Models in which $\omega$ approaches $-1$ from above are known as \enquote{freezing} quintessence \cite{Ferreira:1997hj,Copeland:1997et,Liddle:1998xm,Dutta:2011ik}, and scenarios that start from $\omega=-1$ and evolve to larger values are known as \enquote{thawing} quintessence.

In analogy to primordial inflation one typically imposes the so called slow roll conditions \cite{Chiba:2009sj}
\begin{align}
    \varepsilon_V \equiv \frac{M_\text{Pl.}}{16 \pi}\left(\frac{\frac{\partial V(\theta)}{\partial \theta }}{f_a V(\theta)}\right)^2  = \frac{M_\text{Pl.}^2}{16 \pi f_a^2} \frac{\sin{(\theta)}}{\left(1-\cos{(\theta)}\right)^2} &\overset{!}{\ll} 1,\\
    \eta_V \equiv \left| \frac{M_\text{Pl.}^2}{8 \pi }\frac{\frac{\partial^2 V(\theta)}{\partial \theta^2 }}{f_a^2 V(\theta)} \right|= \left|\frac{M_\text{Pl.}^2}{8 \pi f_a^2} \frac{\cos{(\theta)}}{1-\cos{(\theta)}}\right| &\overset{!}{\ll} 1,
\end{align}
that ensure that the kinetic energy of the axion is subleading compared to its potential, ensuring $\omega<-1/3$.  However for dark energy there exists the additional subtlety, when compared to inflation: There is always  a non-negligible matter component, that also contributes appreciably to the present day expansion of our universe \cite{Linder:2006sv,Scherrer:2007pu}, which is not taken into account in the inflationary slow roll conditions.

Even for inflation, slow roll turns out to be only a sufficient, but not a necessary requirement to generate accelerated expansion \cite{Damour:1997cb,Linde:2001ae}.  The preference for small slow roll parameters in  inflationary cosmology comes from the need of a prolonged inflating phase with a number of e-foldings of  $N_e \gtrsim 50-60$ to solve the horizon and flatness problems.  Since  $N_e \propto 1/\sqrt{\varepsilon_V}$ this implies a small value for $\varepsilon_V$. 
Fast roll models generally lead to a small number of e-folds \cite{Damour:1997cb,Linde:2001ae}. 
However the current phase of accelerated expansion only needs $N_e \gtrsim 1$ so in principle one could have fast roll quintessence.

\subsection{Harmonic regime}
For an initial angle of  $\theta_i \ll 1$ one can approximate the potential as  quadratic, which is known as the harmonic approximation. The conventional choice of initial conditions is $\dot{\theta_i}=0$ and the axion dominates the energy density of the universe today as long as\footnote{We use the $\simeq$ sign due to the non-negligible matter contribution.} 
\begin{align}\label{eq:Friedmann}
    \frac{m_a^2 f_a^2 \theta^2}{2} \simeq \frac{3}{8 \pi} H_0^2 M_\text{Pl.}^2,
\end{align}
where $H_0$ is the present day Hubble rate. Throughout this we employ $M_\text{Pl.} = 1/\sqrt{G_N}=\SI{1.22e19}{\giga\electronvolt}$, where $G_N$ is Newton's gravitational constant.
One finds for the slow roll parameters that
\begin{align}
    \varepsilon_V \simeq \eta_V \simeq \frac{M_\text{Pl.}^2}{4\pi f_a^2 \theta_i^2}.
\end{align}
Slow roll axion quintessence with a quadratic potential can be well fit with the CPL-parameterization 
for redshifts $z\lesssim 1$ \cite{Scherrer:2007pu}. The slow roll conditions are only compatible with $f_a \gg M_\text{PL.}$ \cite{Scherrer:2007pu}.
Arguments from string theory constructions such as the strong version of the Weak Gravity Conjecture \cite{Arkani-Hamed:2006emk} or the Swampland Distance Conjecture \cite{Ooguri:2006in} suggest that the decay constant should be below the reduced Planck scale
\begin{align}\label{eq:WGC}
    f_a < \frac{M_\text{Pl.}}{\sqrt{8\pi}}.
\end{align}
Furthermore even for the case of pure Einstein gravity, there will be contributions from gravitational instantons to the axion potential, that become unsuppressed for $f_a > M_\text{Pl.}/\sqrt{8\pi}$ spoiling its flatness \cite{Alvey:2020nyh}. 
Larger effective   decay constants may arise from two or more aligned axions \cite{Kim:2004rp} with subplanckian decay constants as in clockwork models \cite{Choi:2015fiu,Kaplan:2015fuy}.
However quantum gravitational effects are expected to invalidate these transplanckian field ranges \cite{Montero:2015ofa}. The only viable exception seem to be models with a spectrum of $N$ (almost) degenerate subplanckian axions, whose coherent motion can be described as a single axion with an effective decay constant  $\sqrt{N} f_a >M_\text{Pl.}/\sqrt{8\pi}$ \cite{Kaloper:2005aj}.
Dangerously large $f_a$ can be avoided by abandoning the first slow roll condition (or the slow roll approximation altogether), as we will see in the next section.

\subsection{Hilltop regime}\label{sec:hilltop}
For initial misalignment angles close to the maximum of the cosine-potential at $\theta =\pi$, the harmonic approximation breaks down and the growth of the quadratic potential is tamed by a negative quartic contribution
\begin{align}
    V(\theta) = \frac{m_a^2 f_a^2}{2}\left(\theta^2 -\frac{\theta^4}{12} + \mathcal{O}(\theta^6)\right).
\end{align}
The quadratic term pulls the axion to the minimum at $\theta=0$, whereas the quartic term pulls it up to the top of the potential.
At the hilltop $\theta =\pi$ the energy density for dark energy has to satisfy
\begin{align}\label{eq:Fried}
    2 m_a^2 f_a^2 \simeq \frac{3}{8 \pi} H_0^2 M_\text{Pl.}^2,
\end{align}
and the slow roll parameters are found to be 
\begin{align}
    \varepsilon_V &\simeq 0,\\
    \eta_V &\simeq  \frac{M_\text{Pl.}^2}{16 \pi f_a^2} \label{eq:tachyon},
\end{align}
which indicates that the first slow roll condition is automatically satisfied. The second slow roll condition is satisfied for $f_a > 0.14 \; M_\text{Pl.}$, which leaves some room for subplanckian $f_a$.
Reference \cite{Dutta:2008qn} determined that thawing quintessence starting from the maximum of the potential is possible while violating the second slow roll condition. Their analytical results show that the CPL-parameterization is not a good fit for the hilltop regime, as the equation of state is not linear in the scale factor but rather depends on its cube, unless one takes $\eta_V \rightarrow 0$.

In Ref.~\cite{Kaloper:2005aj} the authors found that $f_a<M_\text{Pl.}/\sqrt{8\pi}$ can be realized for slow roll hilltop quintessence, as long as the initial misalignment angle is tuned exceptionally close to the maximum. The underlying reason is that   at the maximum the axion mass becomes tachyonic $\partial^2 V(\theta)/\partial \theta^2 <0$ (see also eq.~\eqref{eq:tachyon}), implying that this position is unstable. This indicates an exponentially growing displacement $\delta \theta \equiv \pi-\theta \simeq \text{exp}(-m_a t)$ from the maximum, which has to be compensated by the initial condition $\theta_i$, since the axion should remain close to the top and not start rolling before roughly today $t\simeq 1/ H_0$. Using this together with \eqref{eq:Fried} implies the tuning \cite{Kaloper:2005aj}
\begin{align}\label{eq:tuning}
    \delta \theta_i = \pi -\theta_i < e^{- \sqrt{\frac{3}{16\pi}}\frac{M_\text{Pl.}}{f_a}}.
\end{align}
Numerically this corresponds to $\delta \theta_i \simeq 9\times 10^{-2}\;(8\times 10^{-107})$ for $f_a = 10^{18}\;\text{GeV}\;(10^{16}\;\text{GeV})$. This tuning is exacerbated by the fact that inflationary dynamics will delocalize the axion field from its classical value: Quantum fluctuations during inflation with a Hubble rate $H_I< f_a$ lead to spatial inhomogeneities in the axion field. This can be described by a probability distribution for the field centered around its classical value $\braket{\theta}=\theta_i$ with a dispersion due to the fluctuations \cite{Bunch:1978yq}
\begin{align}
    \delta \theta_\text{fluct.}\equiv \sqrt{\braket{\theta^2}} \simeq 
      \begin{cases}
      \frac{H_I}{2\pi f_a}  &\quad \text{for}\quad m_a \ll \sqrt{\varepsilon_I} H_I,\\
      \sqrt{\frac{4}{3}}\pi   &\quad \text{for}\quad   \sqrt{\varepsilon_I} H_I \ll m_a \ll H_I.
    \end{cases}
\end{align}
The first condition $m_a \ll \sqrt{\varepsilon_I} H_I$ with the inflation's slow roll parameter $\varepsilon_I \ll 1$ encodes an axion that is practically massless during inflation and the resulting fluctuation is smaller than $1/(2\pi)$.
Here the second condition ensures, that there are enough inflationary e-foldings \cite{Gorbunov:2011zzc} to drive the axion fluctuation to the largest value  possible for compact scalar fields of $\sqrt{4/3}\pi$ \cite{Chakraborty:2023eoq}.
Imposing the condition in eq.~\eqref{eq:tuning} would exclude the second regime. In the first regime one obtains a limit on $H_I/f_a$, however we refrain from going into detail, since $H_I$ only has an upper limit of $H_I < 6\times 10^{13}\;\text{GeV}$ \cite{BICEP:2021xfz} for canonical, single field, slow roll inflation and might be arbitrarily small compared to the $f_a \lesssim M_\text{Pl.}$. 

When it comes to fitting the current data, one attempt was undertaken by
Ref.~\cite{Tada:2024znt}, who fitted the equation of state and noted that the large value of the CPL-parameter $\omega_a$ seems to require 
a mild violation of both slow roll conditions. The central values for the CPL parameterization $(\omega_0,\omega_a)\simeq (-0.7,-1)$ can be reproduced for a single axion with \cite{Tada:2024znt}\footnote{Note that the authors of \cite{Tada:2024znt} use the potential $1+\text{cos}(\theta)$ with an initial angle of $0.55$, which  corresponds to $\theta_i=\pi -0.55$ for our choice of $1-\text{cos}(\theta)$ and furthermore they normalized $f_a$ to the reduced Planck mass $M_\text{Pl.}/\sqrt{8\pi}$.}
\begin{align}\label{eq:THAW1}
    \frac{m_a}{H_0} =2.95, \quad \frac{f_a}{M_\text{Pl.}}= 0.082, \quad \theta_i =\pi -0.55, \quad \dot{\theta}_i =0,
\end{align}
that rolled down its potential. Moreover the required $f_a$ is safely below the reduced Planck scale. In Ref.~\cite{Bhattacharya:2024kxp} the central values of a Markov Chain Monte Carlo  fit to the datasets from  \verb|Planck|  \cite{Planck:2018vyg,Planck:2019nip},   \verb|Pantheon+|  \cite{Scolnic:2021amr} and  \verb|DESI BAO| data \cite{DESI:2024mwx} were found to be\footnote{Again we converted their result for $f_a$ from reduced Planck units to our choice of $M_\text{Pl.}$.}
\begin{align}\label{eq:THAW2}
    \frac{f_a}{M_\text{Pl.}}= 0.243, \quad \theta_i =2.25, \quad \dot{\theta}_i =0,
\end{align}
but no constraint on $m_a$ was specified. Note that  for both results the required angle is not in the harmonic regime and sufficiently far away from the hilltop at $\theta_i=\pi$ to avoid the previously  discussed fine-tuning problems.

\subsection{Oscillating dark energy}\label{sec:osc}
An oscillating scalar field in a potential, that is convex at the minimum, but concave away from the minimum, could lead to an equation of state averaged over one oscillation period of $\braket{\omega} <-1/3$.
This ideas was proposed for inflation in \cite{Damour:1997cb} and applied to quintessence in \cite{Dutta:2008px}.
The basic picture is, that during one oscillation the field spends enough time close to the hilltop with a large enough potential energy to drive the expansion, and that over many oscillations enough time accumulates, to generate sufficent e-folds.
If one expands the canonical axion potential to fourth order, one finds the equation of state as a function of the initial misalignment angle \cite{Norton:2020ert}
\begin{align}
    \braket{\omega}\simeq -\frac{\theta_i^2}{32},
\end{align}
which holds for $\theta_i<1$, and a numerical calculation reveals that $\braket{\omega} \rightarrow -1$ for $\theta_i\rightarrow \pi$. 
However for oscillating fields with $\braket{\omega}<0$ the authors of Ref.~\cite{Johnson:2008se} observed dynamical  instabilities:
The axion zero mode starts to exponentially excite higher momentum perturbations   via the parametric resonance effect as a consequence of its oscillating effective mass (see  eq.~\eqref{eq:effmass}). Therefore the system is no longer described by a homogeneous and isotropic condensate, due to the spatially varying perturbations. Such an instability would also affect the growth of large scale structure. Ref.~\cite{Johnson:2008se} concludes that oscillating scalar fields, which dominate the energy budget, are not viable  energy candidates.\footnote{The authors of Ref.~\cite{Johnson:2008se} limited their analysis to nearly harmonic potentials and did not consider a cosine potential. Additionally it might occur, that oscillating quintessence is stable only over the timescale it takes, to inflate the universe by $\mathcal{O}(1)$ e-folds.}
To avoid the regime of oscillations, which would require $t_\text{osc} \simeq 1/m_a \ll t_0 \simeq 1/H_0$ to generate many oscillations on cosmological time-scales, we impose that
\begin{align}
    m_a < 10~H_0.
\end{align}
One loophole is that the oscillations could be preceded by a rolling phase and start late enough at e.g. a scale factor of $R=0.8$ to avoid problems with large scale structure \cite{PhysRevD.85.103510}.

\subsection{Initial velocity: Interpolating between Freezing and Thawing}\label{sec:Kick}
Now we relax the assumption $\dot{\theta_i}=0$. The impact of a non-vanishing axion velocity in dark matter models was investigated by \cite{Co:2019jts,Chang:2019tvx} and further studied in \cite{Barman:2021rdr}. 
Suppose the axion starts at an initial time $t_i$ from an initial position $\theta_i$ with  an initial velocity $\dot{\theta_i}>0$\footnote{For a discussion on the sign of the velocity see below eq.~\eqref{eq:table}. In section \ref{sec:BOTH} we show that the cosmological data prefers  $\dot{\theta_i}>0$.}. Then it will proceed to climb up the hill of its potential,  until it reaches a maximal angle $\theta_\text{max} $, before turning around and rolling towards its minimum at $\theta=0$. Using the exact solution to the equation of motion in the harmonic regime during matter radiation with $R\sim t^{2/3}$ and solving $\dot{\theta}(t_\text{max})=0$ we find\footnote{If the axion potential is negligible at this time, we obtain a similar result to the \enquote{weak kinetic misalignment} regime of Refs.~\cite{Chang:2019tvx,Barman:2021rdr}, where $m_a$ is  replaced by $3 H(t_\text{max})/2$.} 
\begin{align}\label{eq:Max}
    \theta_\text{max} \simeq \theta_i+ \left(\frac{R_i}{R_\text{max}}\right)^\frac{3}{2} \frac{\dot{\theta_i}}{m_a},
\end{align}
where $R_\text{max}$ is the scale factor at the time 
\begin{align}
    t_\text{max}\simeq t_i + \frac{1.43 \pi}{m_a},
\end{align}
when the field reaches $\theta_\text{max}$. 
Thus the maximum misalignment angle depends on the initial velocity of the axion and the  time or redshift $z_i$, at which  the axion velocity is turned on. 

As the axion field roll upwards in its potential ($t< t_\text{max}$) it will eventually act as freezing quintessence, once $\omega$ decreases below $-1/3$ and approaches $-1$ (see the middle panel of figure \ref{fig:EOSexamples}).  After it has reached the peak of its trajectory ($t= t_\text{max}$), during which momentarily $\omega=-1$, it starts to roll
down ($t>t_\text{max}$) and will act as thawing quintessence, until $\omega$ grows above $-1/3$. The pivot between both behaviors is essentially governed by $m_a$ since $1/m_a \simeq 1/H_0 \gg t_i$.
A schematic picture of this evolution was depicted in figure \ref{fig:diagram}.
Owing to the dependence shown in eq.~\eqref{eq:Max}
the equation of state for $\theta_i=0,\; \dot{\theta_i}\neq 0$ will be independent of $f_a$. A model independent analytical treatment of quintessence that interpolates between the freezing and thawing behavior was presented in    \cite{Dutta:2011ik,Swaney:2014kpa}. Analytical results for quintessence with an initial kination era were presented in Ref.~\cite{Andriot:2024sif}.

For $\dot{\theta_i}> 2 m_a$ the initial kinetic energy density of the axion $\dot{\theta_i^2}f_a^2/2$ becomes larger than the potential barrier of the cosine potential of $2 m_a^2 f_a^2$. If $\dot{\theta_i}$ is very large\footnote{Unlike the compact field range $\theta\in [-\pi,\pi]$ the velocity $\dot{\theta}$ is is a priori unbounded.}, then the axion is able to roll over the maximum of its periodic potential and explore the neighboring minima. The precise limit on $\dot{\theta_i}$, for this to occur, does not just depend on $m_a$, but also on $f_a$, as the axion can contribute significantly to the background evolution, which will affect the Hubble friction and thus the field excursion. In a toy example with $\Omega_m =0.3,\; \theta_i=0,\; m_a=H_0$ and $f_a=0.1\;M_\text{Pl.}$ we find numerically that $\dot{\theta_i}>120\; m_a$ would be needed to traverse beyond the first period of the axion potential. 

A  axion zero mode rolling over many minima is prone to exciting perturbations of higher momentum modes  via the parametric resonance effect \cite{Jaeckel:2016qjp,Berges:2019dgr,Fonseca:2019ypl,Fonseca:2019lmc,Morgante:2021bks,Eroncel:2022vjg} due to its oscillating effective mass (see  eqns.~\eqref{eq:perturb}-\eqref{eq:effmass} and the discussion in section~\ref{sec:adiab}) and the description in terms of a coherent condensate can break down. This observation is not a \enquote{no-go} theorem for quintessence that previously rolled through many minima, but in any case a dedicated analysis would be required to treat this regime. 
However our best fit points in section \ref{sec:BOTH} feature  small initial velocities of $\dot{\theta_i}= \mathcal{O}(1) \;m_a$ and do not traverse multiple minima of its potential (evident from the lower panel of figure \ref{fig:adiab}), so that  our setup is free from the aforementioned complications.

\section{Sources for the axion velocity}\label{sec:models}
Due to the conversation of the associated Noether charge\footnote{In models with a dynamical radial mode (evolving $f_a$) the redshifting is different, e.g. $\dot{\theta}\sim 1/R$ for $f_a\sim 1/R$.}
\begin{align}
    n_\theta = \dot{\theta} f_a^2
\end{align}
one finds that the axion velocity redshifts as  
\begin{align}
    \dot{\theta}\sim 1/R^3.
\end{align}
Since the scale factor of the universe grows exponentially during inflation, the velocity from the  initial conditions  is expected to be diluted away and typically one sets $\dot{\theta_i}=0$ for quintessence (see Ref.~\cite{Wolf:2024stt} for a counterexample).

One  example for quintessence with a velocity, that causes it to run uphill was proposed in Ref.~\cite{Csaki:2005vq}.  Here the quintessence field rolls down an asymmetric potential that is initially very steep, before it enters a region of the potential at small redshifts that is very shallow and linear in the quintessence field. The velocity is sourced by the mismatch of the potential's slopes. While the authors of \cite{Csaki:2005vq} argue that the overall smallness of such a potential might be a consequence of hidden sector supersymmetry breaking, no mechanism for its required asymmetric shape is specified. For a compact field such as the one considered in this work, one could mimic such an asymmetric potential if the axion decay constant is dynamical and grows significantly at small redshifts in order to flatten the potential. 

However interactions of the axion with other fields such as  derivative couplings to fermions, scalar interactions or topological couplings to gauge fields can also  kick the axion into motion.
This very idea was used in Ref.~\cite{Sakstein:2019fmf}, to dynamically set the initial misalignment angle of a dark matter axion from the dynamics of Baryogenesis or Leptogenesis via a derivative coupling to the baryon number or lepton number current.

If any of these mechanisms are supposed to happen in the very early universe  (e.g. well  before electroweak symmetry breaking), then a very large initial velocity  along the lines of the scheme in Ref.~\cite{Co:2019jts} would be required, in order to have a non-negligible effect at late times of $z=\mathcal{O}(1)$, and to not just shift the axion angle at early times. In this case the axion would traverse many of the neighboring minima of its periodic potential, which, as explained in the previous section \ref{sec:Kick}, comes with the caveat of the axion potentially fragmenting into higher momentum excitations loosing its homogeneity \cite{Jaeckel:2016qjp,Berges:2019dgr,Fonseca:2019ypl,Fonseca:2019lmc,Morgante:2021bks,Eroncel:2022vjg}. Hence we consider only mechanisms that dynamically generate a velocity field for the axion in the late universe, shortly before the epoch of dark energy domination. Thus we need a source term that is relevant in the late universe during matter domination, say e.g. at redshifts $z\leq 3$.

On top of that we need to ensure that the axion kinetic energy is never larger than the dominant component driving the background expansion at the time of the kick, because this kinetic energy has to be sourced from somewhere to begin with (e.g. from a converting fraction of dark matter or dark radiation). This means that
\begin{align}\label{eq:MatDom}
    \frac{\dot{\theta_i}^2 f_a^2}{2} < \rho_\text{mat}(z_i),
\end{align}
which automatically avoids an epoch of kination between matter domination and the epoch of accelerated expansion. If a fraction $c_\text{frac}<1$ of $\rho_\text{mat}(z_i)$ is converted into axion kinetic energy we find that
\begin{align}
    \left|\frac{\dot{\theta}_i}{m_a}\right| = 9.38 \;c_\text{frac}  \left(\frac{1+z_i}{4}\right)^3 \left(\frac{\Omega_m}{0.3}\right) \left(\frac{H_0}{m_a}\right) \left(\frac{M_\text{Pl.}}{f_a}\right).
\end{align}
It is important to stress, that we will mostly concern ourselves with the case, where the bulk of the axion energy density is stored in its potential, and the velocity acts only as a small perturbation, thus the change in the matter density, and therefore $c_\text{frac}$,  will be rather small (see the discussion below eq.~\eqref{eq:source} and in sections \ref{sec:FREE}, \ref{sec:BOTH}).

In the following we present  mechanisms for kicking the axion into motion based on the interactions with  scalars, vectors or fermions.
In appendix \ref{sec:ooE} we demonstrate that $CP$-violating out-of-equilibrium decays of the cosmic neutrino background or fermionic dark matter  would in general produce a far too small velocity of $\dot{\theta_i}/m_a= \mathcal{O}(10^{-33})$. Appendix \ref{sec:HelGauge} shows that a helical background of dark abelian gauge bosons, produced at late times from e.g. dark matter decay,  could lead to larger values of  $\dot{\theta_i}/m_a\lesssim\mathcal{O}(1\%)$.
The only viable case, apart from an asymmetric potential \cite{Csaki:2005vq}, is the coupling to scalars discussed in the next section \ref{sec:AD}.

\subsection{Affleck Dine mechanism or multiple PNGBs}\label{sec:AD}
The Affleck-Dine mechanism \cite{Affleck:1984fy,Dine:1995kz} for the generation of $\dot{\theta_i}$ was applied to PNGBs in Refs.~\cite{Co:2019wyp,Co:2019jts}.
The basic idea is that $\theta$ is the phase of a complex scalar whose vev dynamically breaks the global $\text{U}(1)_\text{X}$
\begin{align}
    \varphi=\frac{r+ f_a}{\sqrt{2}} e^{i \theta}.
\end{align}
The $\text{U}(1)_\text{X}$ is also explicitly broken by operators of the form
\begin{align}
    \frac{\varphi^n}{M_\text{pl.}^{n-4}} + \text{h.c.},
\end{align}
which can act as both a source for the axion potential in eq.~\eqref{eq:Va}, as well as its motion \cite{Berbig:2023uzs}:
If the radial mode $r$  oscillates with an initial amplitude $r_i \geq f_a$ the above operator will convert a part of its motion to a torque in the angular direction. $\dot{\theta}$ will first increase as $R^2$ \cite{Gouttenoire:2021jhk} before reaching an attractor with \cite{Co:2020jtv}
\begin{align}
    \braket{\dot{\theta}} = m_r (r_i),
\end{align}
in terms of the, possibly field dependent, mass $m_r (r_i)$ of the radial mode.
Due to the oscillating driving force from the radial mode, the angular velocity will oscillate around its average value $\braket{\dot{\theta}}$ \cite{Gouttenoire:2021jhk}, until the radial mode is thermalized by additional interactions \cite{Co:2020dya}. After the radial mode has relaxed to its true vacuum $r = f_a$, the angular velocity starts to redshift as $\dot{\theta}\sim 1/R^3$.

However since we expect $f_a$ not to be too far below the Planck scale, these oscillations of $r$  can only have taken place in the very early universe, and not at the time-scale relevant for dark energy. The only way for these dynamics to affect the late universe, is by giving the axion a very large kick. However in that case we expect the axion to potentially traverse many of its minima, and we argued at the end of the previous section \ref{sec:Kick}, why this could be problematic due to fragmentation of the homogeneous axion field.

To sidestep this issue one can imagine a case, where a second PNGB called $A$, that constitutes a component of dark matter, develops a velocity via the Affleck-Dine mechanism at earlier times, and much later transmits it to $a$.
Two PNGBs could communicate  \cite{Kim:2004rp,Choi:2015fiu,Kaplan:2015fuy} via e.g. the potential 
\begin{align}\label{eq:lvlcross}
V_\text{mix} = \Lambda^4 \cos{\left(c_a \frac{a}{f_a}+ c_A \frac{A}{f_A}\right)}.  
\end{align}
Reference  \cite{Domcke:2022wpb} found that such a charge transfer can be very efficient. This could solve the \enquote{Why now?} problem of quintessence models by using e.g. a hidden gauge interaction that confines at the redshift $z_i\leq 10$ to generate $V_\text{mix}$. It is important to ensure that the shift-symmetry breaking couplings to the other PNGB do not increase the dark energy's mass scale. Due to the inherent model-dependence of this approach we refrain from estimating the corresponding $\dot{\theta}_i/ m_a$. 

Note that two-field models with an adiabatically relaxing radial mode (instead of an oscillating one) and large angular velocities were proposed some time ago under the moniker of \enquote{spintessence} \cite{Boyle:2001du,Kasuya:2001pr,Gu:2001tr,Arbey:2001qi,Li:2001xaa,Chiueh:2001ri}, however these setups are apparently plagued by the production of $Q$-balls \cite{Kasuya:2001pr}, even though exceptions exist \cite{Li:2001xaa}. 

Alternatively a coherently oscillating $A$ could source the observed dark matter abundance and its oscillation could kick the axion into motion (see e.g. Ref.~\cite{Cyncynates:2021xzw} for dark matter from multiple axions). For quintessence this was studied recently in Ref.~\cite{Muursepp:2024mbb}, where it was assumed that the quintessence axion $a$ gets its mass from late confinement of a non-abelian gauge group. By using the temperature dependence of this  mass during an ongoing dark sector phase transition, one then obtains the correct equation of state for dark energy.  The resulting energy density of $\rho_a$  turns out to be insufficient to drive the expansion, but once the coupling in eq.~\eqref{eq:lvlcross} is switched on, the level crossing between $a$ and $A$ can drastically enhance $\rho_a$.  However the treatment of this effect is complicated by the fact, that the conversion between the two PNGBs takes place in the deeply non-adiabatic regime \cite{Muursepp:2024mbb}. Our scenario is distinct in the sense that we assume a constant axion mass and, that the majority of the dark energy density comes from its potential instead of  the interplay with dark matter.

Recently another very similar scenario  was investigated in Ref.~\cite{Aboubrahim:2024cyk}, where it was found that the the coupling to dark matter can change the equation of state for dark energy from thawing to a freezing behavior at redshifts  $3\leq z \leq13$, which could be detectable in future surveys. In  the aforementioned work no fit to the currently available data was undertaken, and the setup is also different from our scenario: In Ref.~\cite{Aboubrahim:2024cyk} the evolution of the quintessence field is modified after it has already started rolling (transition from thawing to freezing), whereas we focus on the case where a velocity is injected before it begins to roll (transition from freezing to thawing).

\section{Fit to data}\label{sec:num}
We proceed by fitting the evolution of the cosmic background,  including an initial  velocity $\dot{\theta}_i$ for the axion.
To do so, we employ the compressed  \verb|Planck|  \cite{Planck:2018vyg,Planck:2019nip} and \verb|ACT| \cite{ACT:2023dou,ACT:2023kun} likelihoods for the CMB observables together with the  compressed  \verb|Pantheon+|  \cite{Scolnic:2021amr}  likelihoods for the Supernovae observables, all taken  from table 2 in appendix B of Ref.~\cite{Wolf:2024eph},  as well as the \verb|DESI BAO| data \cite{DESI:2024mwx}.
In \cite{Wolf:2024eph} it was shown that the compressed data reproduces the full likelihoods well, and that the posterior distribution for the CPL parameters recovers the constraints obtained from fitting to the full data. This approach was then used to constrain quadratic quintessence \cite{Wolf:2024eph} and models with non-minimal couplings to gravity \cite{Wolf:2024stt}.
We minimize a $\chi^2$ defined as 
\begin{align}
    \chi^2 =  \left(\mathcal{O}_\text{data}-\mathcal{O}_\theta\right)^t \text{Cov.}^{-1} \left(\mathcal{O}_\text{data}-\mathcal{O}_\theta\right),
\end{align}
where for a given observable $\mathcal{O}$ we denote the data as $\mathcal{O}_\text{data}$,  and $\mathcal{O}_\theta$  is the prediction of our model. Here we introduce the inverse of the covariance matrix $ \text{Cov.}^{-1}$ and a sum over all data sets is understood. 

To assess the results of our fits we compute the reduced $\chi^2$ in terms of  the number of data points $N$ and the number of fitted parameters $k$ 
\begin{align}
    \chi_\text{red}^2 = \frac{\chi^2}{N-k}.
\end{align}
We   further employ  the Akaike Information Criterion (AIC) and the Bayesian Information Criterion (BIC)  \cite{Liddle:2007fy,Akaike1974ANL,1978AnSta...6..461S}
\begin{align}
    \text{AIC} &= 2 k + \chi^2,\\
    \text{BIC} &= k \text{ln}(N) + \chi^2.
\end{align}
For comparison with $\Lambda$CDM we define
\begin{align}
    \Delta \chi^2 &\equiv \chi^2_\text{$\Lambda$CDM}- \chi^2,\\
    \Delta \text{AIC} &\equiv \text{AIC}_\text{$\Lambda$CDM}-  \text{AIC},\\
    \Delta \text{BIC} &\equiv \text{BIC}_\text{$\Lambda$CDM}-\text{BIC}.
\end{align}

After determining the set of microscopic parameters that give the best fit, we compute the corresponding CPL parameters for the sake of illustration  following Ref.~\cite{Wolf:2024eph} by minimizing
\begin{align}\label{eq:chiCPL}
    \chi^2_\text{CPL} = \left(\frac{\mathcal{O}_\text{data}}{\mathcal{O}_\theta} \left(\mathcal{O}_\text{CPL}-\mathcal{O}_\theta\right) \right)^t \text{Cov.}^{-1} \frac{\mathcal{O}_\text{data}}{\mathcal{O}_\theta}\left(\mathcal{O}_\text{CPL}-\mathcal{O}_\theta\right),
\end{align}
where $\mathcal{O}_\text{CPL}$ is the prediction computed in the CPL model with the Hubble rate given by eq.~\eqref{eq:HubCPL}. The factors of $\mathcal{O}_\text{data}/\mathcal{O}_\theta$ make sure that the relative error for each dataset is used instead of the absolute errors \cite{Wolf:2024eph}.


\begin{figure}
    \centering
    \includegraphics[width=0.6\textwidth]{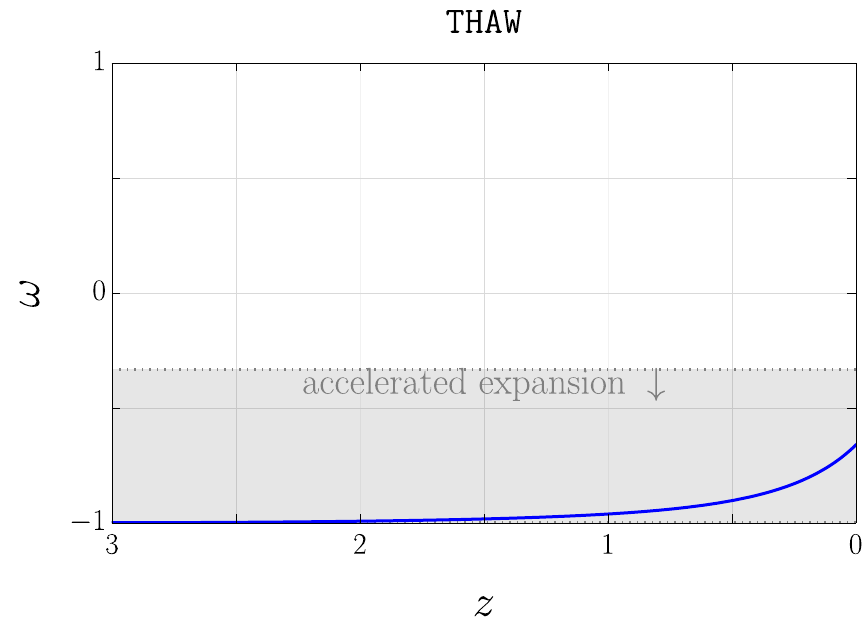}  
    \includegraphics[width=0.6\textwidth]{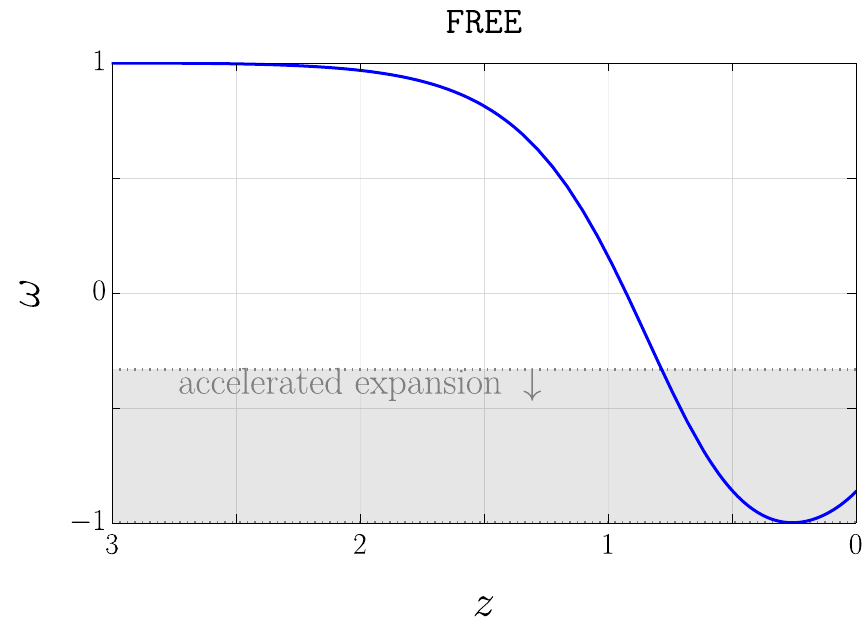}
    \includegraphics[width=0.6\textwidth]{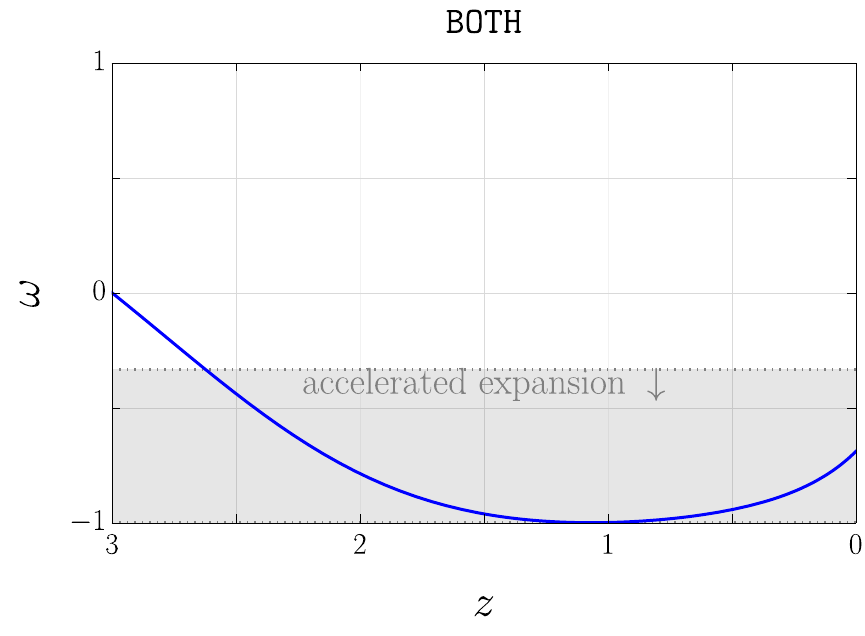}  
    \caption{Examples for the evolution of the equation of state as a function of redshift $z$, where the parameters $m_a=H_0,\;f_a=0.1\; M_\text{Pl.},\;\Omega_m=0.3,\;z_i=3$ where chosen for illustration only and we consider \textit{(top)} $\theta_i=0.1,\; \dot{\theta_i}/m_a =0$ , \textit{(middle)} $\theta_i=0,\; \dot{\theta_i}/m_a =0.1$  and \textit{(bottom)} $\theta_i=0.1,\; \dot{\theta_i}/m_a =0.1$. Accelerated expansion occurs in the gray shaded area, where $\omega \leq -1/3$.}
    \label{fig:EOSexamples}
\end{figure}

\subsection{Numerical Treatment}
We recast the Klein-Gordon equation for the axion  and the Friedmann equation for the Hubble rate $H$ (instead of the scale factor $R$) in terms of redshift $z$ (instead of cosmic time $t$), where dashes denote derivatives with respect to $z$
\begin{align}
    \theta''(z) + \left(\frac{H'(z)}{H(z)}-\frac{2}{1+z}\right) \theta'(z) + \frac{m_a^2}{(1+z)^2 H(z)^2} \sin(\theta(z))&=0\label{eq:ThetaEvolve},\\
    \frac{H'(z)}{H_0} - \frac{3 H_0}{2 (1+z) H(z)} \left((1+z)^3 \Omega_m  +  \frac{ (1+z)^2 H(z)^2 f_a^2}{\rho_\text{crit}} \theta'(z)^2 \right)&=0 \label{eq:HubbleEvolve}.
\end{align}
The initial conditions are 
\begin{align}
    \theta(z_i) =\theta_i,\quad \theta'(z_i) = -\frac{\dot{\theta}_i}{(1+z_i) H(z_i)}, \quad H(z_i) = H_0 \sqrt{\Omega_m (1+z_i)^3   + \frac{\rho_\theta(z_i)}{\rho_\text{crit}}}\label{eq:Initial}.
\end{align}
Through out this work we assume the absence of spatial curvature $\Omega_K =0$ implying 
\begin{align}\label{eq:table}
    \Omega_\theta + \Omega_m =1, \quad \text{with} \quad \Omega_\theta \equiv \frac{\rho_\theta(z=0)}{\rho_\text{crit}}.
\end{align}
In the following we focus on three scenarios:
\begin{itemize}
    \item \texttt{THAW} defined as $\theta_i\neq 0,\;\dot{\theta}_i=0$
    \item \texttt{FREE} defined as $\theta_i= 0,\;\dot{\theta}_i \neq 0$
    \item \texttt{BOTH} defined as $\theta_i\neq 0,\;\dot{\theta}_i\neq0$
\end{itemize}
In figure \ref{fig:EOSexamples} we depicted the evolution of the equation of state computed from eqs.~\eqref{eq:ThetaEvolve}-\eqref{eq:Initial} for these three cases.   
One can see, that all three scenarios produce very similar equations of state at late redshifts $z<1$, but differ significantly at earlier times. This can be understood by noting that the freezing benchmark \texttt{FREE}, that starts with only kinetic energy from the kick,  begins with  $\omega=1$. Conversely the mixed scenario \texttt{BOTH} with both potential and kinetic energy at the initial time starts with $-1 < \omega <0$, which itself is distinguishable from the thawing scenario \texttt{THAW}, whose initial equation of state is $\omega=-1$.

In the case of $\dot{\theta_i}=0$ we can choose the domain of $\theta_i$ as $[0,\pi]$, due to the reflection symmetry of the potential in eq.~\eqref{eq:Va}. The same reasoning for the case of $\theta_i =0$  implies that we can limit ourselves to  $\dot{\theta_i}>0$. If both initial conditions are non-vanishing we can still work with  $\theta_i \in [0,\pi]$, but we have to keep track of the sign of $\dot{\theta_i}$ as the axion could get kicked upwards $(\dot{\theta_i}>0)$ or downwards $(\dot{\theta_i}<0)$. 

\subsection{CMB}
Since the axion velocity generated at $z_i$ has to be sourced from some other energy density, we allow the matter density $\Omega_m$ to vary between the early times $z\simeq 1090 \gg z_i$ of CMB decoupling and the late times $z<3\leq z_i$ of the SN and BAO observations. In practice that means, that we evaluate the CMB quantities in terms of $\Omega_m^\text{rec}$, which is related to the present day matter density $\Omega_m$ by 
\begin{align}\label{eq:source}
    \Omega_m^\text{rec} = \Omega_m + \frac{\dot{\theta}_i^2 f_a^2}{2\rho_\text{crit} (1+z_i)^3}.
\end{align}
We will see in section \ref{sec:BOTH}, that we  only encounter small shifts of the matter density parameter $\Omega_m^\text{rec}- \Omega_m \simeq \mathcal{O}(1\%)$, so we barely modify the predictions for the CMB.
The sound horizon is defined in terms of sound speed  $c_s$ computed from the baryon (photon) energy density $\rho_B\;(\rho_\gamma)$
\begin{align}\label{eq:rd}
    r_d(z) = \int_{z_d}^\infty  \text{d}z \frac{c_s(z)}{H(z)}, \quad \text{with} \quad  c_s(z) = \frac{1}{3\sqrt{\left(1+ \frac{3\rho_B(z)}{4\rho_\gamma(z)}\right)}},
\end{align}
 and we parameterize these quantities following Ref.~\cite{Schoneberg:2021uak}
\begin{align}
  \frac{3\rho_B(z)}{4\rho_\gamma(z)} = \frac{667.2}{1+z} \left(\frac{\Omega_b h^2}{0.022}\right)^2 \left(\frac{\SI{2.7225}{\kelvin}}{T_0}\right)^4,
\end{align}
where $T_0$ is the present day CMB temperature.
The Hubble rate at $z>z_i$ is given by
\begin{align}\label{eq:hubearly}
    H(z) = H_0 \sqrt{\Omega_m^\text{rec} (1+z)^3 + \Omega_r (1+z)^4 + \frac{V_\theta(z)}{\rho_\text{crit}}},
\end{align}
and the radiation energy density  reads \cite{Schoneberg:2021uak}
\begin{align}
     \Omega_r = \frac{2.47\times 10^{-5}}{h^2}\left(1+\frac{7}{8}\left(\frac{4}{11}\right)^\frac{4}{3}N_\text{eff.}\right),
\end{align}
where $N_\text{eff.}$ is the contribution due to neutrinos. Throughout this work we fix
\begin{align}\label{eq:constants}
   T_0 = \SI{2.7225}{\kelvin}, \quad \Omega_b h^2 = 0.02235, \quad N_\text{eff.} = 3.04,  
\end{align}
because our setup does not modify the early time cosmology. 
The relevant quantities for the fit are the sound horizon's angular scale \cite{WMAP:2008lyn,Planck:2015bue,Chen:2018dbv}
\begin{align}
    l_A = \pi (1+z_*) \frac{D_A(z_*)}{r_d(z_*)}
\end{align}
as well as the CMB shift parameter \cite{WMAP:2008lyn,Planck:2015bue,Chen:2018dbv},
\begin{align}
    R(z_*) = (1+z_*) \sqrt{\Omega_m^\text{rec} H_0^2} D_A(z_*),
\end{align}
that together encode the position of the first acoustic peak in the CMB temperature power spectrum \cite{Elgaroy:2007bv}.
Both quantities depend on the angular diameter distance
\begin{align}
    D_A(z)= \frac{1}{1+z}\int_0^z \frac{\text{d}z'}{H(z')},
\end{align}
and the redshift at the time of last scattering $z_*\simeq 1090$, for which we use the following fitting function from Ref.~\cite{Hu:1995en}
\begin{align}
    z_* &= 1048 \left(1+ 0.00124 \left(\Omega_b h^2\right)^{-0.738}\right)\left(1+ g_1\left(\Omega_m^\text{rec} h^2\right)^{g_2}\right),\\
    g_1 & = \frac{0.0783 \left(\Omega_b h^2\right)^{-0.238}}{1+ 39.5 \left(\Omega_b h^2\right)^{0.763}},\\
    g_2 &= \frac{0.56}{1+21.1 \left(\Omega_b h^2\right)^{1.81}}.
\end{align}
For the computation of $D_A(z_*)$ we split the integral in two regions: For $z_i< z \leq z_*$ we use the analytical formula for the Hubble rate in eq.~\eqref{eq:hubearly} and for $0<z\leq z_i$ we use the solution for $H(z)/H_0$ obtained from the numerical solution of eq.~\eqref{eq:HubbleEvolve}, as the energy density in radiation has redshifted away to negligible amounts.
The compressed data for $l_A$ and $R(z_*)$ obtained from  \verb|Planck|  \cite{Planck:2018vyg,Planck:2019nip} and \verb|ACT| \cite{ACT:2023dou,ACT:2023kun} can be found in table 2 of appendix B in Ref.~\cite{Wolf:2024eph}.

\subsection{DESI BAO data}
\verb|DESI| BAO \cite{DESI:2024mwx} measured  the quantity
\begin{align}
    D_M(z) = \int_0^z \frac{\text{d}z'}{H(z')},
\end{align}
and the equivalent distance
\begin{align}
    D_H(z) = \frac{1}{H(z)},
\end{align}
for redshifts between 0.51 and 2.33, as well as the angle-averaged combination 
\begin{align}
    D_V(z) = \left(z D_M(z) D_H(z)\right)^\frac{1}{3}
\end{align}
for $z=0.295$ and $z=1.49$. Since \verb|DESI| BAO data is only sensitive to $D_{M,H,V}/r_d(z_d)\sim 1/(H_0 r_d(z_d))$ in terms of the sound horizon $r_d(z_d)$ defined in eq. \eqref{eq:rd}, one needs CMB data to break this degeneracy and extract $H_0$. Here  $z_d\simeq 1060$ is the redshift at the time of baryon-photon-decoupling, for which we use the fitting formula from Ref.~\cite{Hu:1995en}
\begin{align}
    z_d &= 1345 \frac{\left(\Omega_m^\text{rec} h^2\right)^{0.251}}{1+0.659 \left(\Omega_m^\text{rec} h^2\right)^{0.828}} \left(1+ b_1\left(\Omega_b h^2\right)^{b_2}\right),\\
    b_1 &= \frac{0.313}{\left(\Omega_m^\text{rec} h^2\right)^{0.419}} \left(1+0.607 \left(\Omega_m^\text{rec} h^2\right)^{0.674}\right),\\
    b_2 &= 0.238 \left(\Omega_m^\text{rec} h^2\right)^{0.223},
\end{align}
in terms of $\Omega_m^\text{rec}$ defined in eq.~\eqref{eq:source}. 
We use the data for $D_{M,H,V}/r_d(z_d)$ from Ref.~\cite{DESI:2024mwx}.

\subsection{Supernovae}\label{sec:SN}
The supernova data involves the comoving distance
\begin{align}
    D_L(z) = (1+z) \int_0^z \frac{\text{d}z'}{H(z')},
\end{align}
and in Ref.~\cite{Riess:2017lxs} it was shown that the data can be compressed into measuring $H_0/H(z)$. In Ref.~\cite{Wolf:2024eph} this procedure  was carried out for the \verb|Pantheon+| \cite{Scolnic:2021amr} data set, up to a redshift of 2 (see table 2 of appendix B in \cite{Wolf:2024eph}).
The author of Ref.~\cite{Efstathiou:2024xcq} found that there seems to be a statistical error in the \verb|DES5Y| dataset that leads to a shift of the magnitudes compared to the \verb|Pantheon+| SN compilation, which reduces the significance of the time evolving dark energy from $3.9 \sigma$ found by the \verb|DESI| collaboration \cite{DESI:2024mwx} down to $2.5\sigma$ (see also Refs.~\cite{Colgain:2024ksa,Colgain:2024mtg} for analyses with different conclusions).\footnote{After the completion of this work the \texttt{DES} collaboration reanalyzed the apparent discrepancy between the SN datasets and found that it reduces the significance for time dependent dark energy only down to  $3.3 \sigma$ \cite{DES:2025tir}.} Hence we only include  \verb|Pantheon+|  in our analysis. 

\section{Discussion}\label{sec:discuss}
\begin{figure}[t]
    \centering
    \includegraphics[width=0.95\textwidth]{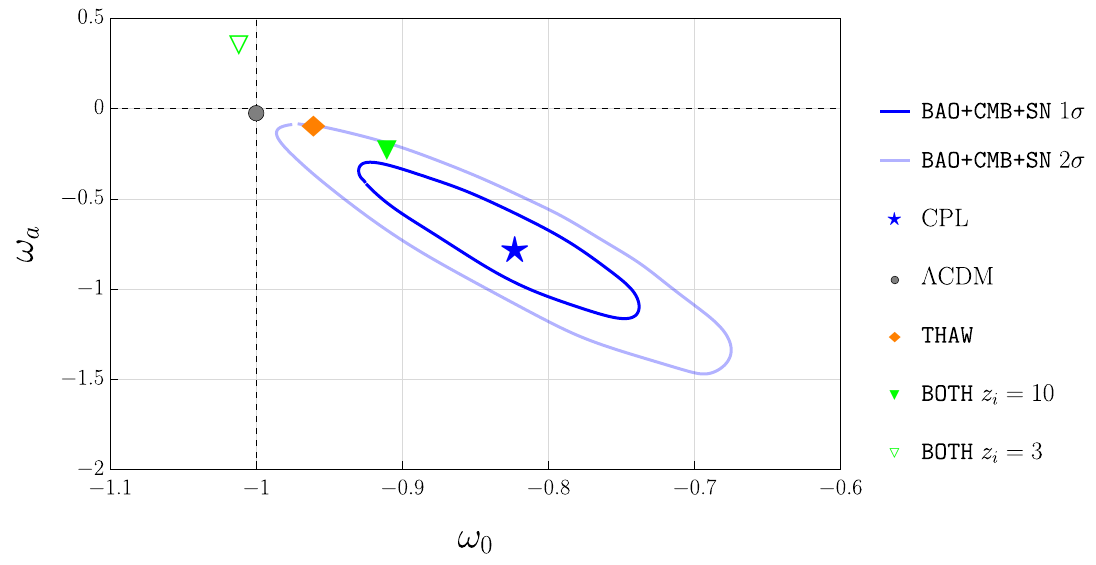}   
    \cprotect\caption{Parameter space of the CPL model when compared to a combination of  CMB  data
    from \verb|Planck|  \cite{Planck:2018vyg,Planck:2019nip} and \verb|ACT| \cite{ACT:2023dou,ACT:2023kun}, SN data from   \verb|Pantheon+|  \cite{Scolnic:2021amr}  and \verb|DESI| BAO data \cite{DESI:2024mwx}.
    The blue star showcases our best fit result for the CPL cosmology (see eq.~\eqref{eq:HubCPL}) from table \ref{tab:CDM}. The $1\sigma$ and $2\sigma$ contours for the combined fit to datasets where digitized from Ref.~\cite{Wolf:2024eph}. The black dot corresponds to conventional $\Lambda$CDM. We further show the CPL parameters obtained from fitting the phenomenologically viable best fit points for the axion scenarios in table \ref{tab:model1} to the CPL parameterization via minimization of  eq.~\eqref{eq:chiCPL}.}
    \label{fig:CPL}
\end{figure}

\cprotect\subsection{$\Lambda$CDM and CPL}
We validate our fitting routine by first constraining $\Lambda$CDM and the CPL dark energy model.
The results were summarized in the table~\ref{tab:CDM} and for both scenarios we obtain similar  results for $\Omega_m$ and
\begin{align}
    h \equiv \frac{H_0}{100 \frac{\text{km}}{\text{s}\;\text{Mpc}}}.
\end{align}
Ref.~\cite{DESI:2024mwx} found that the  CPL model can not alleviate the $H_0$-tension with the \verb|SH0ES|  result \cite{Riess:2021jrx}, which is based on a distance-ladder calibrated with Cepheids, beyond a residual tension of $(2-3)\sigma$, depending on the combination of datasets used.  Refs.~\cite{Banerjee:2020xcn,Lee:2022cyh} analyzed the interplay of dark energy and the $H_0$-tension and found that $\omega_0>-1$ leads to a decrease in $H_0$.

In general we find that the fit to CPL (see eq.~\eqref{eq:HubCPL}) has a slightly lower $\chi^2$ of about 27, when compared to the value of around 37 obtained for $\Lambda$CDM, and the best fit points of $\omega_0 = -0.823, \; \omega_a = -0.783$ agree well with the central values in eq.~\eqref{eq:DESIlimits} from the analysis of Ref.~\cite{DESI:2024mwx}. 
The $\chi^2$ values obtained for each dataset were collected in table \ref{tab:CDM2}.
For illustration we plot the best fit points in the $\omega_a$ versus $\omega_0$ plane together with the $1\sigma$ and $2\sigma$ contours that we digitized from Ref.~\cite{Wolf:2024eph} in figure \ref{fig:CPL}. One can see that our best fit point for CPL agrees very well with the central value of the $1\sigma$ and $2\sigma$ ellipses.

We fit $N=21$ data points and $\Lambda$CDM and CPL have $k=0$ \cite{Wolf:2024eph} and $k=2$ free parameters respectively, because one parameter is always fixed by the Friedmann equation.
The CPL model has $\Delta \chi^2= 9.23$ and $\Delta \text{AIC}=5.523,\; \Delta \text{BIC} = 3.434$. All three measures agree on CPL being a better fit to the data. We find $\chi_\text{red}^2=1.591$, which is close to unity. Thus our analysis reproduces the conclusion from Ref.~\cite{DESI:2024mwx} that dynamical dark energy in the CPL model is slightly preferred over $\Lambda$CDM.

\begin{table}[t]
\centering
\begin{tabular}{|c||c|c|c||c|c|c|} 
\hline
 & $z_i$ & $\omega_0$& $\omega_a$ & $h$ & $\Omega_m$& $\chi^2$\\
\hline
\hline
 $\;\;\;\Lambda$CDM & - & -1 & 0 & 0.678 & 0.312 & 36.568 \\
\hline 
 CPL & - & -0.823 & -0.783 & 0.682 & 0.310 & 27.045\\ 
\hline
\hline
\texttt{THAW} & 10& -0.960 & -0.085 & 0.673 & 0.316 & 0.002\\
\hline
\multirow{2}{0.1\linewidth}{$\;\;\;$\texttt{BOTH}}& 3 & -1.012& 0.361 & 0.683 & 0.284 & 256.782\\
& 10 & -0.910 & -0.218 & 0.683 & 0.298 & 47.582\\
\hline
\end{tabular}
\cprotect\caption{The first two lines contain the best fit points for the $\Lambda$CDM and CPL (see eq.~\eqref{eq:HubCPL}) cosmologies. The last two lines showcase the CPL values obtained by minimizing eq.~\eqref{eq:chiCPL} for the best fit parameters of the axion scenarios \texttt{THAW} and \texttt{BOTH}, that can be found in table \ref{tab:model1}.}
\label{tab:CDM}
\end{table}

\begin{table}[t]
\centering
\begin{tabular}{|c||c|c|c||c|c|c|c|} 
\hline
 & $\chi^2_\text{BAO}$& $\chi^2_\text{CMB}$ & $\chi^2_\text{SN}$ &$\chi^2_\text{red}$ & $\Delta \chi^2$&  $\Delta$AIC & $\Delta$BIC\\
\hline
\hline
 $\;\;\;\Lambda$CDM & 20.062 & 1.538 & 14.968 & 1.741 &  0 & 0& 0 \\
\hline 
 CPL & 12.642 & 0.112 & 14.290 & 1.591  & 9.523 & 5.523 & 3.434\\ 
\hline
\end{tabular}
\cprotect\caption{$\chi^2$-values for the individual datasets for the best fit points depicted in table \ref{tab:CDM} for the  $\Lambda$CDM and CPL cosmologies.}
\label{tab:CDM2}
\end{table}

\cprotect\subsection{\texttt{THAW} $(\theta_i \neq 0,\; \dot{\theta}_i =0)$}\label{sec:THAW}
\begin{figure}
    \centering
    \hspace{-0.1\textwidth}
    \includegraphics[width=0.73\textwidth]{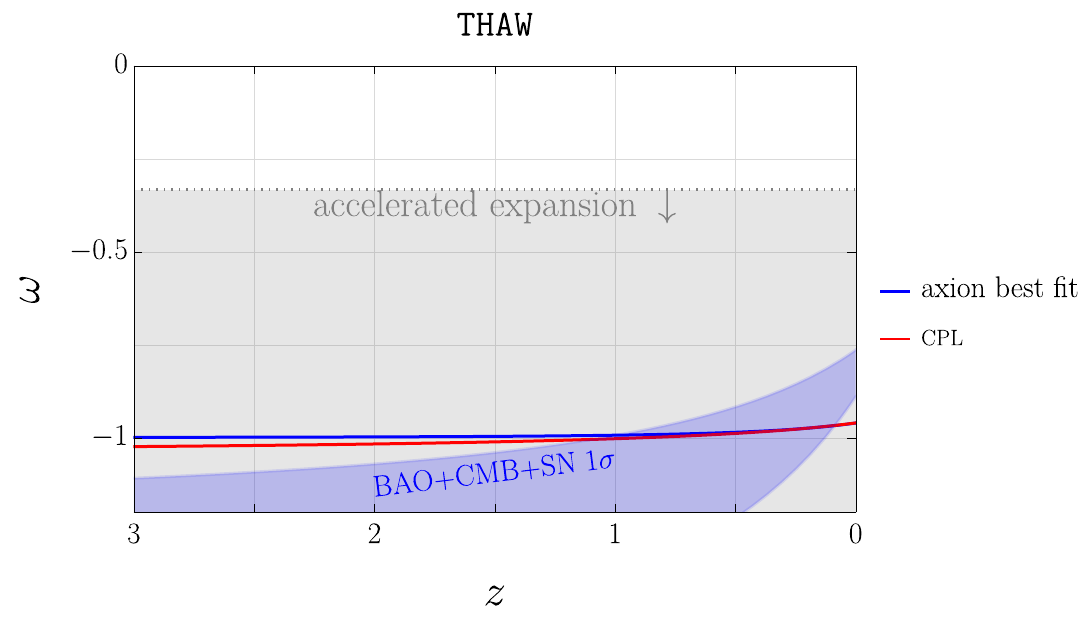}   
    \includegraphics[width=0.83\textwidth]{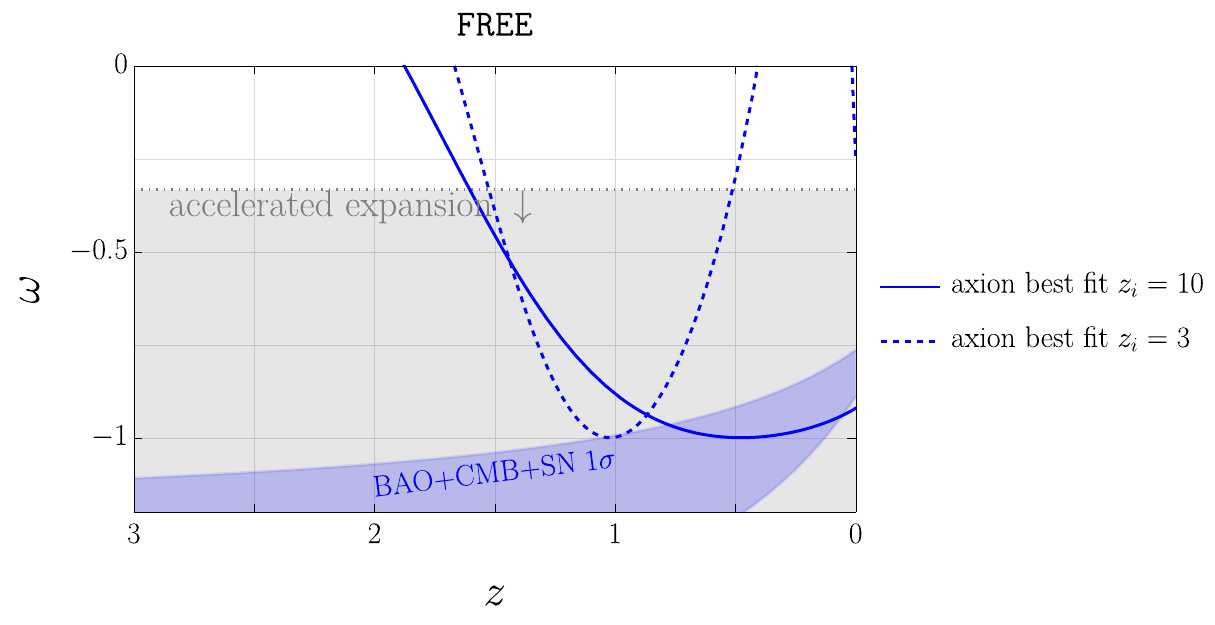}
    \includegraphics[width=0.83\textwidth]{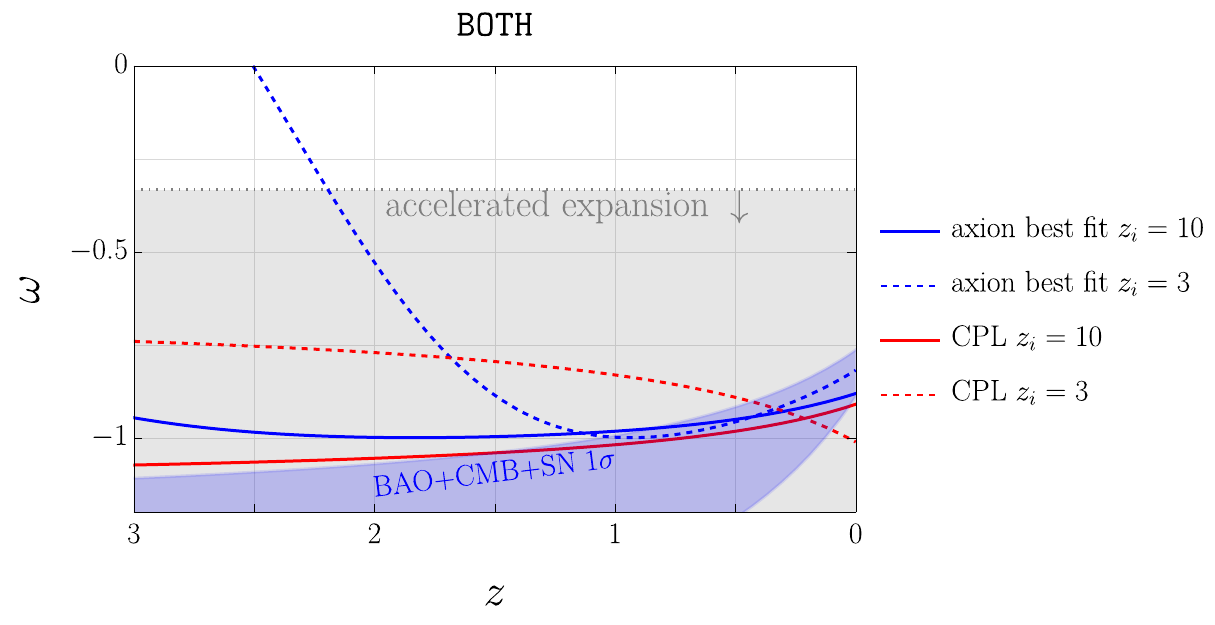} 
    \cprotect\caption{Equation of state for the best fit points of the axion model collected in table \ref{tab:model1}, depicted in blue, together with the equation of state for the corresponding CPL parameters from table \ref{tab:CDM}, depicted in red as functions of redshift $z$. The nomenclature of the three scenarios can be found below eq.~\eqref{eq:table}.
    The $1\sigma$ contours from the fit to BAO, CMB and SN data were computed from eq.~\eqref{eq:EOSCPL} with the parameters in eq.~\eqref{eq:DESIlimits}.}
    \label{fig:EOS-ALL}
\end{figure}
The results for the thawing axion quintessence scenario \texttt{THAW} can be found in table \ref{tab:model1} and the contributions from the individual datasets are shown in table \ref{tab:model2}. We find a  decay constant of $f_a=0.133 \; M_\text{Pl.}$ and an axion mass of $m_a= 1.618\; H_0$ together with an initial angle $\theta_i =2.606$, that is not in the harmonic regime, but also sufficiently small compared to $\pi$, which avoids the finetuning 
of slow roll axion quintessence in the hilltop regime discussed in section \ref{sec:hilltop}. The axion begins to roll at $z_\text{roll}\simeq 8.9$.
Order of magnitude wise our best fit points agree with the results of Ref.~\cite{Tada:2024znt} in eq.~\eqref{eq:THAW1}. The $\mathcal{O}(1)$ differences arise, because we fit the combined data with quantities computed directly from our numerical solution for $H(z)$, whereas the authors of \cite{Tada:2024znt} fit the CPL equation of state, which does not depend on $h$, choosing $\Omega_m =0.3$ and  $(\omega_0,\omega_a)\simeq (-0.7,-1)$. The central values of the full Markov Chain Monte Carlo fit of Ref.~\cite{Bhattacharya:2024kxp} in eq.~\eqref{eq:THAW2} also agree with our findings at the order of magnitude level and the differences can be attributed to the different analysis strategies together with the priors used in \cite{Bhattacharya:2024kxp}. 

It is evident from table \ref{tab:CDM}, that \texttt{THAW} prefers a value of $h=0.673$ for the Hubble rate today, which is slightly smaller than the result for $\Lambda$CDM of $h=0.678$.

For the thawing scenario we find that  $\Delta\chi^2 = 1.602$  and $\Delta \text{AIC}=-2.398,\; \Delta \text{BIC}=-4.487$. Only the first measure prefers thawing quintessence over $\Lambda$CDM, while the other two disfavor it. This mirrors the findings of Ref.~\cite{Wolf:2024eph}, where similar conclusions were reached for the case of thawing hilltop quintessence with a quadratic potential. For illustration we compute the CPL parameters for our best fit point by minimizing eq.~\eqref{eq:chiCPL}. The result was tabulated in table \ref{tab:CDM} and depicted in figure \ref{fig:CPL}, where one can see that the parameter point lies on the $2\sigma$ contour of the combined fit. Additionally we show the equation of state for the best fit point and the associated CPL parameters in the upper panel of figure \ref{fig:EOS-ALL}. The slopes of the equation of state agree well with each other, with the most noticeable difference being that the CPL case crosses the phantom divide $\omega <-1$ for redshifts $z>1$, whereas the canonical quintessence model with a positive kinetic term stays always above $\omega=-1$. That this can happen, when mapping quintessence models to the CPL parameter space, was already pointed out in \cite{Linder:2006xb,Wolf:2023uno,Shlivko:2024llw}.

\cprotect\subsection{\texttt{FREE} $(\theta_i = 0,\; \dot{\theta}_i \neq 0)$ }\label{sec:FREE}
We find that the case, where the axion starts from $\theta_i=0$ so that all of its   energy stems  from the kick, is not a good description of the data, both for injecting the axion velocity either at $z_i=3$ or $z_i=10$. The results of the fit were tabulated in \ref{tab:model1}.
This is evinced by the very large values of $\chi^2=8325.481$ for $z_i=3$ and $\chi^2=8325.481$ for $z_i=10$ as well as the $\chi^2_\text{red}$ and $\Delta\text{AIC},\;\Delta\text{BIC}$ collected in table \ref{tab:model2}. These parameter points are clearly not viable, as the Hubble rate today $h=0.471$ is about $30\%$ too small and the matter density is found to be $\Omega_m \simeq 1$. This arises due to the fact, that the axion energy density is tiny as a consequence of the small values of $\dot{\theta_i} = 1.217\times 10^{-4} m_a$ for $z_i=3$ and $f_a = 8.33\times 10^{-4} M_\text{Pl.}$ for $z_i=10$. As  $\Omega_m \simeq 1$ implies $\Omega_\theta\simeq 0$ from the closure relation in eq.~\eqref{eq:table}, we would have no accelerated expansion   today for these parameter points.

From table \ref{tab:model2} we also deduce that the large $\chi^2$ values are mainly driven by the CMB, and we checked that this dataset is responsible for the  pull towards $h=0.471$ and $\Omega_m\simeq1$. The most likely culprit for this finding is our modeling of the source of the axion velocity injection by a shift of the early time matter density parameter defined in eq.~\eqref{eq:source}.
We verified. that without this term  the aforementioned behavior would disappear.  A physical way to understand this observation is, that the CMB data forces $\Omega_m^\text{rec}$ to be very close to $\Omega_m$, and thus the term $\dot{\theta_i^2} f_a^2 / (2\rho_\text{crit})$ has to be much smaller than unity. Since the total axion energy density parameter today for the \texttt{FREE} scenario has to be smaller than $\dot{\theta_i^2} f_a^2/ (2\rho_\text{crit})$ due to redshfiting, we find that $\Omega_\theta\simeq 0$ which explains the preference for $\Omega_m\simeq 1$. This conclusion can only be avoided, if we have non-vanishing initial potential energy, which is why we consider $\theta_i\neq0$ in the next section \ref{sec:BOTH}.
For completeness we show the equation of state for the best fit parameters in the middle panel of figure \ref{fig:EOS-ALL}, but we do not consider this scenario further.

\cprotect\subsection{\texttt{BOTH} $(\theta_i \neq 0,\; \dot{\theta}_i \neq0)$}\label{sec:BOTH}
Our last scenario has both non-vanishing $\theta_i$ and $\dot{\theta_i}$. We chose to inject the velocity at $z_i\geq 3$, because the earliest redshift probed by our combination of BAO and SN surveys is $z=2.33$.
The best fit parameters can be found in table \ref{tab:model1}: For $z_i=10$ we need an initial velocity of $\dot{\theta_i} = 4.010 \;m_a$, and if the velocity is injected at a later time $z_i=3$ a smaller value of $\dot{\theta_i} = 0.591\; m_a$ is required, because the field velocity has less time to be diluted by redshifting.
The smaller velocity needed for $z_i=3$ might be realized in the dark photon model of appendix \ref{sec:HelGauge}, if we chose a large value of $\Delta N_A$ (see eq.~\eqref{eq:KickGauge} in the appendix), whereas the larger $\dot{\theta_i}$  for $z_i=10$ can only be accommodated with the coupled scalar field models of section \ref{sec:AD}.
We did not impose a sign for the velocity in our fitting procedure and find that data prefers $\dot{\theta_i}>0$.
In the lower panel of figure \ref{fig:adiab} one can observe that the axion only traverses a small angular field range, and never reaches the neighboring minima of its periodic potential. Since the axion begins to move at $z_i=10\;(3)$, we find that it  starts rolling before (after) it would have in the \texttt{THAW} scenario with $z_\text{roll}\simeq 8.9$\footnote{For \texttt{BOTH} with  $z_i=3,\;10$ we find, that, if we were to set $\dot{\theta_i}=0$, both sets of best fit parameters imply that the axion should start to roll at $z_\text{roll}\simeq 8.8$. 
Note that we only numerically solve the equations of motion starting at $z_i$.
For $z_i=10$ this is self-consistent as $z_\text{roll}<z_i$ and the axion gets kicked before it would start to roll. For $z_i=3$ the axion would actually not be frozen by Hubble friction anymore at the time of the kick, since here $z_\text{roll}>z_i$.
One can imagine other sources of friction \cite{Gomes:2023dat,Berghaus:2020ekh,Berghaus:2023ypi,Berghaus:2024kra}, apart from the expansion of the universe, that could keep the axion from rolling before $z=3$. We leave a detailed treatment of this regime for future work.\label{foot:11}}.

As the kicked axion has less time to roll down its potential for $z_i=3$, a smaller initial angle of 
$\theta_i=0.368$ is necessary compared to $\theta_i= 0.849$ for $z_i=10$.
One finds the decay constants $f_a = 0.854\; M_\text{Pl.}\;(0.466 \;M_\text{Pl.})$ for $z_i=3\;(10)$ and 
the larger (smaller) decay constant  for $z_i=3\;(10)$ is needed to fix $\Omega_\theta$ for the smaller (larger) values of $\theta_i$ and $\dot{\theta_i}$. It is evident, that the presence of a non-vanishing $\dot{\theta_i}$ allows for smaller values of $\theta_i$, even farther away from the hilltop than in the \texttt{THAW} scenario.
Overall larger $f_a$ are required compared to the case of \texttt{THAW}.

We find   $\Delta \chi^2 = 10.133\;(10.679)$ and $\Delta\text{AIC}=4.132\;(4.679)$, $\Delta \text{BIC}=0.998\;(1.545)$  for $z_i=3\;(10)$ from table \ref{tab:model2}, demonstrating that for both $z_i$ all three measures prefer these realizations of quintessence with an initial velocity over $\Lambda$CDM. The most likely reason for the improved fit, when compared to the the conventional \texttt{THAW} scenario in section \ref{sec:THAW}, is the additional free parameter $\dot{\theta_i}$.
For $z_i =10$ the value of $\chi_\text{red}^2$ is marginally smaller than for $z_i=3$, and from all previously mentioned measures we deduce, that the parameters for $z_i=10$ are slightly more favored than for $z_i=3$.

To understand why this scenario offers a much better fit to the CMB data compared to the previous case of \texttt{FREE} with $\theta_i=0$, note that early time shift of the matter density parameter in eq.~\eqref{eq:source} is $\Omega_m^\text{red}-\Omega_m = 0.025\;(0.011)$ for $z_i = 3\;(10)$. This shows that for \texttt{BOTH} the bulk of the axion's energy density comes from its potential energy, and that the kick at $z_i$ acts as more of a small perturbation. Thus for $\theta_i\neq 0$ we can have $\Omega_\theta\simeq 0.7$ and $\Omega_m \simeq 0.3$.

When it comes to the Hubble rate today, one can see from table \ref{tab:CDM}, that both realizations of \texttt{BOTH} prefer $h=0.683$, which is slightly larger than the value of $h=0.673$ preferred by \texttt{THAW} and similar to the result of $h=0.682$ for CPL. 

Additionally we compute the CPL parameters that correspond to our best fit points by minimizing eq.~\eqref{eq:chiCPL}, and the CPL parameters can be found in table \ref{tab:CDM} as well as in figure \ref{fig:CPL}. The CPL parameters for \texttt{BOTH} with $z_i=10$ lie on the $1\sigma$ contour and are closer to the best fit CPL value from table \ref{tab:CDM} than the CPL parameter that corresponds to the \texttt{THAW} scenario, which illustrates the improved fit to the data. We plot the evolution of the equation of state in the lowest panel of figure \ref{fig:EOS-ALL}. One can see that the shapes for the quintessence model and CPL agree well, again up to the phantom crossing of the CPL parameterization at redshifts below about $z=1$ (see the discussion in Refs.~\cite{Linder:2006xb,Wolf:2023uno,Shlivko:2024llw}). 

On the other hand for $z_i=3$ we find a much worse fit to obtain the CPL parameters with $\chi^2=256.782$ when compared to $z_i=10$ with $\chi^2=47.582$ (see table \ref{tab:CDM}). The resulting CPL parameters with $\omega_0=-1.012<-1$ would correspond to a phantom equation of state today. When plotted in the $\omega_a$ versus $\omega_0$ plane, one can see that the point for $z_i=3$ is the most distant from the $2\sigma$ contour compared to all other points. This may seem surprising, given the fact that the \texttt{BOTH} quintessence scenario with $z_i=3$ was only marginally less favored than the one for $z_i=10$ (see table \ref{tab:model2}). The most probable cause for this behavior seems to be, that the CPL parameterization is not a good fit for axion quintessence with a substantial amount of kinetic energy injected at late times. This can be observed in the lowest panel of figure \ref{fig:EOS-ALL}, where the dashed lines correspond to $z_i=3$ and it is evident that the steep quintessence behavior is not at all  captured by the corresponding, rather flat CPL curve.

\begin{table}[t]
\centering
\begin{tabular}{ |c||c|c|c|c|c||c|c|c|c||c|| } 
\hline
&$z_i$  &$m_a/H_0$ & $f_a/M_\text{Pl.}$ & $\theta_i$ & $\dot{\theta}_i/m_a$ &  $h$ & $\Omega_m$ & $\chi^2$\\
\hline
\hline
\texttt{THAW} & 10 & 1.618 & 0.133 & 2.606 & 0  & 0.673 & 0.316   & 34.966\\
\hline
\hline
\multirow{2}{0.1\linewidth}{$\;\;\;$\texttt{FREE}}& 3 &4.861 &  0.170 & 0 & $1.217\times 10^{-4}$  & 0.471 & $1-8\times 10^{-10}$   & 8325.481\\
& 10 &1.082 & $8.33\times10^{-4}$ & 0 & 0.204 & 0.471 & $1-5\times 10^{-11}$   & 8696.270\\
\hline 
\hline
\multirow{2}{0.1\linewidth}{$\;\;\;$\texttt{BOTH}} & 3 & 1.232 & 0.854 & 0.368 & 0.591  & 0.683 & 0.284  & 26.436\\
 &10 & 1.021 & 0.466 & 0.849 & 4.010 & 0.683 & 0.298   & 25.889\\
\hline
\end{tabular}
\cprotect\caption{Best fit points for the axion quintessence model for various choices of the initial conditions $\theta_i$ and $\dot{\theta_i}$.} 
\label{tab:model1}
\end{table}

\begin{table}[t]
\centering
\begin{tabular}{ |c||c|c|c|c||c|c|c|c| } 
\hline
 & $z_i$ &  $\chi^2_\text{BAO}$& $\chi^2_\text{CMB}$ & $\chi^2_\text{SN}$ &$\chi^2_\text{red}$ & $\Delta \chi^2$&  $\Delta$AIC & $\Delta$BIC\\
\hline
\hline
\texttt{THAW} & 10& 19.542 & 1.888 & 13.537 & 1.840 &  1.602  & -2.398& -4.487 \\
\hline
\hline
\multirow{2}{0.1\linewidth}{$\;\;\;$\texttt{FREE}} &3  & 2319.850 & 5799.450 & 566.163 &  438.183& -8648.900   & -8292.910 & -8295.000 \\
&10 & 2323.780 & 5806.300 & 566.185 &  457.698 & -8659.700  &  -8663.700& -8665.790 \\
\hline 
\hline
\multirow{2}{0.1\linewidth}{$\;\;\;$\texttt{BOTH}}  & 3 & 12.350 & 0.058 & 14.027 &1.469 & 10.133 & 4.132 & 0.998 \\
& 10 & 12.500 & 0.0002 & 13.387 &  1.438 & 10.679 & 4.679 &1.545 \\
\hline
\end{tabular}
\cprotect\caption{$\chi^2$-values for the individual datasets for the best fit points collected in table \ref{tab:model1} for three axion quintessence scenarios.}
\label{tab:model2}
\end{table}

\section{Additional Constraints}\label{sec:constr}
\subsection{Fragmentation of the kicked axion}\label{sec:adiab}
\begin{figure}
    \centering
    \includegraphics[width=0.9\textwidth]{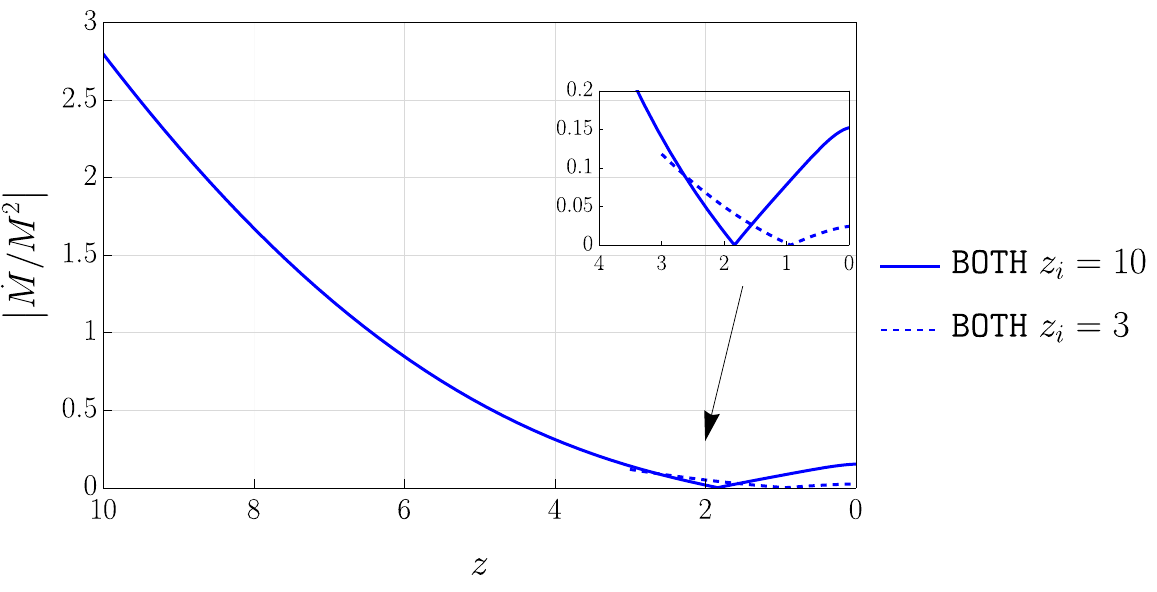}
    \includegraphics[width=0.9\textwidth]{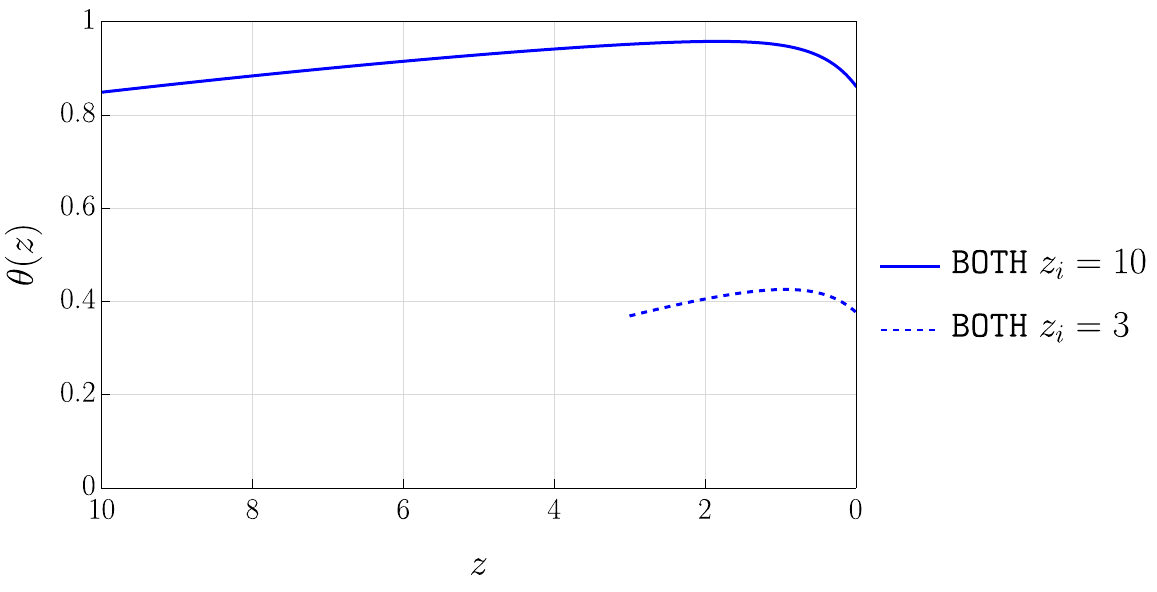}
    \cprotect\caption{\textit{(top)} Adiabaticity parameter defined in eq.~\eqref{eq:adiab} as a function of redshift for the best fit points of the \texttt{BOTH} scenario from table \ref{tab:model1}. \textit{(bottom)} Evolution of the dimensionless axion field $\theta= a/f_a$ as a function of redshift for the aforementioned parameters.}
    \label{fig:adiab}
\end{figure}

The axion is not the only dark energy candidate that can be self-consistently coupled to matter: Chameleon fields  \cite{Khoury:2003aq,Khoury:2003rn}  with non-minimal couplings to the trace of the stress energy tensor  can experience \enquote{kicks} during radiation domination, because a fermion  turning non-relativistic breaks conformal symmetry \cite{Erickcek:2013oma,Erickcek:2013dea,Sakstein:2019fmf}. When the chameleon field is kicked from its potential minimum, runs up the hill and turns around (completely analogous to the dynamics described in section~\ref{sec:Kick}), its effective mass can change drastically, leading to a violation of the adiabaticity condition \cite{Gorbunov:2011zzc,Kolb:2023ydq}
\begin{align}
    \left|\frac{\dot{M}}{M^2}\right| \ll 1
\end{align}
and thus the production of higher momentum modes via the parametric resonance effect. The back-reaction of these excitations on the zero-mode acts as a dissipation term strongly affecting the evolution of the zero-mode \cite{Erickcek:2013oma,Erickcek:2013dea}, so the classical description breaks down. For our case of axion dark energy the equation of motion for these excited modes $\delta \theta_k$ of  momentum $k$ reads \cite{Jaeckel:2016qjp,Berges:2019dgr,Fonseca:2019ypl,Fonseca:2019lmc,Morgante:2021bks,Eroncel:2022vjg}
\begin{align}\label{eq:perturb}
    \ddot{\delta \theta_k} + 3 H \dot{\delta \theta_k} + \left(\frac{k^2}{R^2} +  m_a^2 \cos{(\theta)}\right) \delta \theta_k =0,
\end{align}
and hence we define the $\theta$-dependent effective mass
\begin{align}\label{eq:effmass}
    M^2 \equiv \frac{1}{f_a^2} \frac{\partial^2 V(\theta)}{\partial \theta^2} = m_a^2 \cos{(\theta)},
\end{align}
which allows us to express the adiabaticity condition as 
\begin{align}\label{eq:adiab}
     \left|\frac{\dot{M}}{M^2}\right| =  \left|\frac{\dot{\theta} (z)\tan{(\theta(z))}}{2 m_a\sqrt{\cos{(\theta(z))}}}\right| \ll 1.
\end{align}
The upper panel of figure \ref{fig:adiab} shows the evolution of the adiabaticity parameter for the best fit parameters of the  \texttt{BOTH}  scenario from table \ref{tab:model1}. For $z_i=10$ with $\dot{\theta_i} = 4.010 \;m_a$ we obtain that  $|\dot{M}(z_i)/M(z_i)|^2\simeq 2.7$ implying that the adiabaticity is mildly violated initially. For $z_i=3$ with the smaller  $\dot{\theta_i} = 0.591\; m_a$ the initial adiabaticity parameter is only $|\dot{M}(z_i)/M(z_i)
|^2\simeq 0.12$ and this can be understood from the fact that equation \eqref{eq:adiab} is linear in $\dot{\theta}/m_a$. The corresponding evolution of the axion field was depicted in the lower panel of figure \ref{fig:adiab}.
For both benchmarks we find that the adiabaticity violation decreases as the axion climbs up its potential well, and reaches zero once the axion has reached its maximal angle at $z \simeq 0.9 \;(1.8)$ for $z_i = 3\;(10)$. As the axion descends the adiabaticity parameter starts to grow again.

\subsection{String Theory Conjectures}\label{sec:QG}
\begin{figure}[t]
    \centering
    \includegraphics[width=0.9\textwidth]{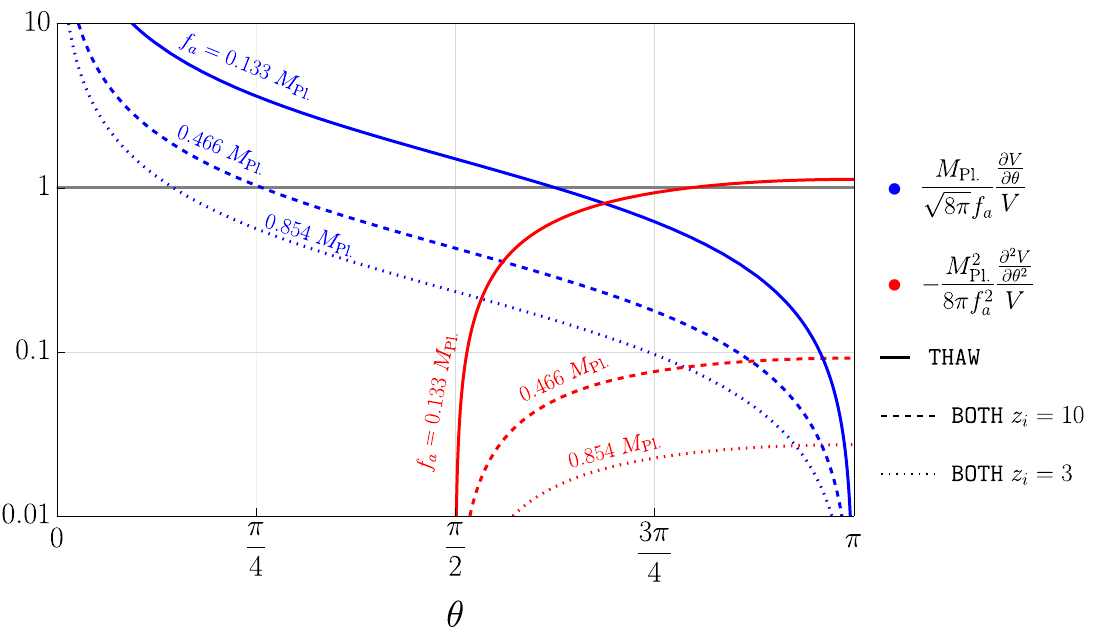}
    \cprotect\caption{Swampland coefficients defined in eq.~\eqref{eq:swamp} as a function of the axion field value for the phenomenologically viable best fit points from table \ref{tab:model1}. Note that for each scenario the quintessential axion only traverses a fraction of the depicted field range.}
    \label{fig:swamp}
\end{figure}

Details on string theory realizations of dark energy and string motivated conjectures can be found in the review \cite{Cicoli:2023opf}.
The decay constants for the best fit points from table \ref{tab:model1} read in units of the reduced Planck mass
\begin{align}\label{eq:fa}
    \frac{\sqrt{8\pi} f_a}{M_\text{Pl.}} = \begin{cases}
       0.667\quad \texttt{THAW},\\
       2.336\quad \texttt{BOTH}\; z_i =3,\\
        4.281\quad \texttt{BOTH}\; z_i =10.
    \end{cases}
\end{align}
One can see that only the thawing hilltop scenario \texttt{THAW} has a subplanckian decay constants, whereas both benchmarks for the \texttt{BOTH} scenario with a late time axion velocity injection have transplanckian decay constants, that violate the bound from the Weak Gravity  \cite{Arkani-Hamed:2006emk} and the Swampland Distance  \cite{Ooguri:2006in}  Conjectures in eq.~\eqref{eq:WGC}. If these conjectures turn out to be true, then this could imply that the \texttt{BOTH} scenario is only viable for models with $N$ axions along the lines of Ref.~\cite{Kaloper:2005aj}. 
The problem with the decay constants in eq.~\eqref{eq:fa} is, that the weak gravity conjecture \cite{Arkani-Hamed:2006emk} demands for axions, that there exists an instanton with an action $S_\text{inst}$ that is bounded by
\begin{align}
    S_\text{inst} \lesssim \frac{M_\text{Pl.}}{\sqrt{8\pi}f_a}.
\end{align}
We used the condition for calculational control over the instanton expansion $S_\text{inst}\gtrsim 1$ to obtain the previous equation \eqref{eq:WGC}. The strong version of the Weak Gravity Conjecture says that this must be the instanton with the smallest action \cite{Bhattacharya:2024kxp}. 
For our scenarios we obtain 
\begin{align}
    S_\text{inst} \lesssim \begin{cases}
       1.499\quad \texttt{THAW},\\
       0.428\quad \texttt{BOTH}\; z_i =3,\\
        0.234\quad \texttt{BOTH}\; z_i =10,
    \end{cases}
\end{align}
which shows, that for the \texttt{BOTH} scenario we are not in the regime of calculational control.
Since the axion potential in eq.~\eqref{eq:Va} is expected to be generated by instantonic effects from confining gauge theories or quantum gravity \cite{Kallosh:1995hi} with an energy scale $\Lambda$, as $m_a^2 f_a^2 \simeq \Lambda^4 e^{-S_\text{inst}}$,  we find that 
\begin{align}
    \Lambda \simeq \SI{3e-3}{\electronvolt} \sqrt{\frac{m_a}{H_0} \frac{f_a}{M_\text{Pl.}}} e^{\frac{S_\text{inst}}{4}}.
\end{align}
Together with the previous rage of $S_\text{inst}$ this implies that $\Lambda$ can not be too far away from the meV scale and the underlying physics might become important for late time cosmology.
Note that we assumed that the axion potential is already present at $z_i=10$, but if it arises from e.g. a confining sector with the above $\Lambda$, it might actually turn at later times.  

For completeness we show that the canonically normalized axion $a=\theta f_a$ traverses the following subplanckian field ranges
\begin{align}\label{eq:ranges}
      \frac{\sqrt{8\pi} \Delta a}{M_\text{Pl.}} = 
      \begin{cases}
        0.133\quad  \texttt{THAW},\\
       0.318\quad  \texttt{BOTH}\; z_i =3,\\
     0.477\quad \texttt{BOTH}\; z_i =10,    
      \end{cases}
\end{align}
and we stress again that the axion never leaves the first minimum of its potential (see e.g. the lower panel in figure \ref{fig:adiab}). The field excursion of $a$ can be small, as the recent universe only needs to inflate with $\mathcal{O}(1)$ e-folds.
For the scenario \texttt{THAW} we just took the difference between $a(z=0)$ and $a(z_i)$, whereas for the scenarios with an initial velocity we first computed the distance from $a(z_i)$ to  $a_\text{max} = \theta_\text{max} f_a$ (see eq.~\eqref{eq:Max} for the definition of $\theta_\text{max}$) and then added the distance from $a_\text{max}$ to $a(z=0)$.

The  de Sitter Conjecture \cite{Obied:2018sgi,Garg:2018reu,Ooguri:2018wrx} is motivated by the fact that it is generally very hard to realize de Sitter vacua in string theory and it demands, that scalar potentials should not be too flat
\begin{align}\label{eq:swamp}
    \frac{M_\text{Pl.}}{\sqrt{8\pi} f_a} \frac{\frac{\partial V}{\partial \theta}}{V} \geq c_1 \quad \text{or} \quad    -\frac{M_\text{Pl.}^2}{8\pi f_a^2} \frac{\frac{\partial^2 V}{\partial \theta^2}}{V} \geq c_2,
\end{align}
where one expects the dimensionless constants $c_{1,2}$ to be of  $\mathcal{O}(1)$.  
Due to the structure of the axion potential in eq.~\eqref{eq:Va} the above Swampland coefficients only depend on $\theta$ and $f_a$. We depicted their dependence on $\theta$  for the phenomenologically viable best fit points from table \ref{tab:model1} in figure \ref{fig:swamp}. This plot evinces that the   Swampland coefficients decrease for larger values of $f_a$. For all three benchmark points there are regions in the axion field space, where both Swampland coefficients are smaller than unity, and thus the de Sitter conjecture is violated. The axion field range, in which the conjecture is violated, increases with increasing $f_a$: For the \texttt{THAW} scenario with $f_a=0.133\;M_\text{Pl.}$ only the region around $\theta\simeq 0.7\pi$ violates the conjecture, whereas for the scenario \texttt{BOTH} with $z_i = 3\;(10)$ and $f_a = 0.854\;M_\text{Pl.}\;(0.466\;M_\text{Pl.})$ the entire range of $\theta > 0.15 \pi\;(0.25 \pi)$ is in violation of the conjecture. 

We conclude that our best fit point for the \texttt{THAW} scenario is the least incompatible with the aforementioned set of conjectures, as it features a subplanckian decay constant and violates the de Sitter conjecture only in a small region of the axion field space. 
The realizations of the \texttt{BOTH} scenario, which are slightly favored by cosmological data over the \texttt{THAW} case, violate the Weak Gravity Conjecture together with the Swampland Distance Conjecture  due to their transplanckian values of $f_a$ and the de Sitter Conjecture as their potentials are too flat.

\subsection{Cosmic Birefringence}
Recently the rotation angle of the CMB polarization was measured to be \cite{Minami:2020odp,Diego-Palazuelos:2022dsq,Eskilt:2022cff}
\begin{align}\label{eq:biref}
    \beta = (0.34\pm 0.01)^\circ,
\end{align}
with a significance of  more than $3\sigma$. This isotropic birefringence can be explained if the two photon helicities propagate differently due to a parity violating background. 
One way to induce this effect is to consider a coupling of the axion to the electromagnetic field strength tensor $\mathcal{F}$
\begin{align}
    \frac{g_{a\gamma\gamma}}{4} a \mathcal{F}_{\mu\nu}\tilde{\mathcal{F}}^{\mu\nu},
\end{align}
which could arise in certain UV completions of our scenario or for instance in the dark photon model of appendix \ref{sec:HelGauge}, if the dark $\text{U}(1)$ has gauge kinetic mixing with hypercharge.
It is important to stress that a priori this coupling exists independently of our scenario for axion quintessence with an initial velocity. The rotation angle is proportional to the axion field excursion after recombination \cite{Carroll:1989vb,Fujita:2020ecn,Lin:2022niw}
\begin{align}
    |\beta|= g_{a\gamma\gamma} \frac{\Delta a}{2},
\end{align}
and using our numerical results for the field excursions of the best fit points in eq.~\eqref{eq:ranges} together with the observed angle in eq.~\eqref{eq:biref} we obtain that
\begin{align}
           g_{a\gamma\gamma} = \frac{1}{M_\text{Pl.}}
      \begin{cases}
        0.45\quad  \texttt{THAW},\\
       0.19\quad  \texttt{BOTH}\; z_i =3,\\
     0.13\quad \texttt{BOTH}\; z_i =10.    
      \end{cases}
\end{align}
While these couplings are so tiny that they evade all known constraints, one could make the argument that if  $g_{a\gamma\gamma}$ was ever determined by other means than the CMB birefringence, one could pinpoint from this (hypothetical) measurement if thawing or kicked quintessence were realized.

\section{Conclusions}\label{sec:conc}
Motivated by recent developments for axion dark matter \cite{Co:2019jts,Chang:2019tvx}  we have  entertained the possibility that PNGB quintessence possesses a non-vanishing field velocity $\dot{\theta_i}$. 

We  argued that sourcing a sufficiently large velocity in the early universe could lead to fragmentation of the axion via the parametric resonance effect \cite{Jaeckel:2016qjp,Berges:2019dgr,Fonseca:2019ypl,Fonseca:2019lmc,Morgante:2021bks,Eroncel:2022vjg}, and chose to generate the velocity at late times instead. This motivates, why we inject the velocity at $z_i=3$ and $z_i=10$, since
the earliest redshift probed by our combination of BAO and SN surveys is $z=2.33$. 
We sketched multiple mechanisms to generate this velocity based on couplings to other scalars, derivative couplings to fermions or a coupling to a background of dark vector bosons, and found that the fermion and dark gauge boson models typically predict  too small velocities.

To remain model independent for our cosmological analysis, we parameterized the source of the velocity injection as arising from a shift in the matter density parameter $\Omega_m$, that is at most larger by $\mathcal{O}(1\%)$ than usual at the time of CMB decoupling (see eq.~\eqref{eq:source}). 

A $\chi^2$-fit to data from  \verb|DESI| BAO  \cite{DESI:2024hhd},    \verb|Planck|  \cite{Planck:2018vyg,Planck:2019nip}, \verb|ACT| \cite{ACT:2023dou,ACT:2023kun} and  \verb|Pantheon+|  \cite{Scolnic:2021amr} along the lines of  Ref.~\cite{Wolf:2024eph}  at the background level revealed that this \enquote{kicked} quintessence with $\dot{\theta_i}=\mathcal{O}(1) m_a$  and $\theta_i \neq 0$  is a better fit to data than both the cosmological concordance model $\Lambda$CDM and the conventional \enquote{thawing} quintessence with $\dot{\theta_i}=0$. The presented model fares better than the conventional thawing quintessence as we fit one additional free parameter $\dot{\theta}_i$. Our numerical results were compiled in the tables \ref{tab:CDM}-\ref{tab:model2}.

Variations of the Affleck-Dine mechanism, a coupling to a dark matter PNGB or an asymmetric potential that changes at late times \cite{Csaki:2005vq} seem to be the most promising ways of generating such values of $\dot{\theta_i}$.
Our results show that the later (earlier) injection at $z_i=3\;(10)$ needs a smaller (larger) velocity and a larger (smaller) decay constant $f_a$. 
Unlike the result for the thawing case, we found transplanckian decay constants in the kicked scenario, which do not agree with the Weak Gravity Conjecture \cite{Arkani-Hamed:2006emk} or the Swampland Distance Conjecture \cite{Ooguri:2006in}, but \textit{might} be accommodated in models with $N$ almost degenerate axions
\cite{Kaloper:2005aj}. 
The aforementioned parameter space also features too shallow potentials, which violate the de Sitter conjecture \cite{Obied:2018sgi,Garg:2018reu,Ooguri:2018wrx}.
Additionally we found that the adiabaticity condition for the absence of axion fragmentation is mildly violated initially for $z_i=10$ due to the larger $\dot{\theta_i}$, but not for $z_i=3$, so smaller $z_i$ appear to be favored.
Thus while choosing larger $z_i$ can help with satisfying the conjectures due to the smaller $f_a$, the violation of adiabaticity will be exacerbated. 

It turns out that the scenario with $\dot{\theta_i}\neq0$ and $\theta_i=0$ is not a good fit to data, because without an initial potential all of  the axion energy has to be sourced by the kick. The small shift in $\Omega_m$ allowed by the CMB data then basically leads to $\Omega_\theta\simeq0$, which would imply the absence of dark energy today.

Our work is only intended to be a preliminary proof of concept, and we leave a thorough statistical analysis together with a complete exploration of the parameter space spanned by $\{m_a,f_a,\theta_i,\dot{\theta_i},z_i\}$ for future investigations. Such a refined analysis should also vary $\Delta N_\text{eff.}$ or the baryon fraction $\Omega_b h^2$ as free parameters (see eq.~\eqref{eq:constants}). Also one should fit the data with $z_i$ as a free parameter instead of fixing it to select benchmarks.

We parameterized the kick by changing the initial conditions for the axion and thus we can only treat the case, where it gets kicked  \textit{before} it would start to roll (see also footnote \ref{foot:11}). It would be worthwhile to further analyze the case where it gets pushed \textit{while already rolling} as in Ref.~\cite{Aboubrahim:2024cyk}, by including the interaction with other fields as a source term in the axion equation of motion.  


Since the axion velocity is sourced by a late time change of the dark matter relic abundance our proposal might lead to observable consequences for structure formation. Furthermore  the axion could possess additional suppressed couplings to SM fields, not required for the quintessence dynamics explored in this paper, that can manifest themselves as time-dependent constants of nature \cite{Carroll:1998zi}.  One such example is a topological coupling to the electromagnetic gauge field from e.g. gauge kinetic mixing between hypercharge and the dark $\text{U}(1)$ in appendix \ref{sec:HelGauge}, which could be responsible for the isotropic cosmic birefringence observed in the \verb|Planck| polarization data \cite{Minami:2020odp,Diego-Palazuelos:2022dsq,Eskilt:2022cff}.

Another important future direction for the proposed scenario is to analyze it at the level of linear perturbations and eventually extend the analysis to a full Markov Chain Monte Carlo simulation.  In order to compute the perturbations we need to focus on a concrete model generating the axion velocity, with couplings to dark matter being the most likely candidate. This would need to involve carefully treating the interplay of dark matter and dark energy  microphysics, since most analyses only use a  phenomenological fluid description  for their interaction (see e.g. \cite{Gavela:2009cy,Ghedini:2024mdu}).

Additionally we anticipate, that the next data release by the \verb|DESI| collaboration in the near future and data from \verb|EUCLID| \cite{2011arXiv1110.3193L} will be instrumental for further testing or even ruling out our proposal. 

\section*{Acknowledgments}
The author would like to thank Miguel Escudero for invaluable discussions during the early stages of this project and Juan Herrero-Garcia for useful feedback on the manuscript. 
Furthermore the author is grateful for the kind hospitality of the Max Planck Instiute for Nuclear Physics Heidelberg, the CERN theory group, as well  the Kavli Institute of Physics and Mathematics of the Universe at the University of Tokyo, where parts of this work were completed. 
MB is supported by \enquote{Consolidación Investigadora Grant CNS2022-135592}, funded also by \enquote{European Union NextGenerationEU/PRTR}.

\appendix 
\section*{\Large{Appendix}}
\section{Late asymmetry generation from neutrino or dark matter decay}\label{sec:ooE}
Consider some fermions $f$ charged under $\text{U}_\text{X}(1)$, that have the following derivative couplings
\begin{align}\label{eq:deriv}
    \frac{\partial_\mu a}{f_a} \sum_f c_f \overline{f} \gamma^\mu f.
\end{align}
A CPT-violating background with $\dot{\theta}\neq0$ 
acts as an external chemical potential for $f$ \cite{Cohen:1987vi,Cohen:1988kt} 
\begin{align}
    \mu_f  = c_f \dot{\theta},
\end{align}
that is related to the particle-antiparticle asymmetry of $f$ at the temperature $T$ via $\Delta n_f = \mu_f T^2$.
The basic idea now is to use a \textit{separate} mechanism to source the $f$-asymmetry $\Delta n_f$, which will generate the
Noether charge 
\begin{align}
 n_\theta = \dot{\theta} f_a^2 \simeq \frac{ \Delta n_f}{c_f},  
\end{align}
and hence an axion velocity $\dot{\theta}$. 
Such an approach was first considered in Ref.~\cite{Alonso:2020com} to dynamically set the initial misalignment angle  for axion dark matter via Baryogenesis or Leptogenesis. 
This idea can be understood as the inverse process of \enquote{Spontaneous Baryogenesis} \cite{Cohen:1987vi,Cohen:1988kt}, and it is complementary to scenarios in which the motion of the quintessence field drives Spontaneous Baryogenesis \cite{Li:2001st,DeFelice:2002ir}. 
Here we note that in the Baryogenesis context $\text{U}_\text{X}(1)$ does not need to be identified with baryon number $B$, but could be related to a variety of possible quantum numbers \cite{Chiba:2003vp,Takahashi:2003db}: lepton number $L$ \cite{Ibe:2015nfa}, $B-L$,  right handed $B+L$ \cite{Maleknejad:2020yys}, right handed neutrino number in Dirac neutrino mass models \cite{Berbig:2023uzs}, right handed electron number (or any accidentally conserved SM fermion number) \cite{Domcke:2020quw}, dark matter number \cite{March-Russell:2011ang} or the  Peccei-Quinn charge \cite{Co:2019wyp}.

For concreteness we consider decaying neutrino dark radiation producing an asymmetry, because we are interested in processes occurring at late times $z=\mathcal{O}(1-10)$ (see the discussion in section \ref{sec:Kick}):
We focus on  neutrino mass generation via the Type I Seesaw mechanism \cite{Minkowski:1977sc,Yanagida:1979as,Gell-Mann:1979vob,Glashow:1979nm, PhysRevLett.44.912}  with $\text{U}_\text{X}(1)$ playing the role of  global lepton number 
\begin{align}
    -\mathcal{L}_\text{UV} = Y_L \overline{L} \tilde{H} \nu_R +  Y_R\; \varphi \overline{\nu_R^c} \nu_R + \text{h.c.},
\end{align}
with the charges  $Q_\text{X}[\nu_R]=Q_\text{X}[L] = 1$ and $\varphi$ being a complex scalar with  $Q_\text{X}[\varphi]=-2$, whose phase is the axion.
Here we denote the $\text{SU}(2)_\text{L}$ conjugate of the SM Higgs doublet as $\tilde{H}$. 
We perform field redefinitions of $\nu_{L,R}$ to move the axion into derivative couplings of the form of eq.~\eqref{eq:deriv} for $\nu_{L,R}$.
Lepton number is broken by the vev $f_a\lesssim M_\text{Pl.}$ of the radial mode of $\varphi$, and the right handed neutrino mass can be parametrically below $f_a$ for small entries of the Yukawa matrix $Y_R$. At   energies below electroweak symmetry breaking (EWSB) one can imagine that additional degrees of freedom facilitate the following effective operator 
\begin{align}\label{eq:decay}
    -\mathcal{L}_\text{IR} = \frac{c_\nu}{\Lambda_\text{UV}} \overline{L}\tilde{H} \sigma \psi \overset{\text{EWSB}}{\rightarrow} Y_\nu \overline{\nu_L}\sigma \psi,\quad Y_\nu \equiv \frac{c_\nu}{\sqrt{2}} \frac{v_H}{\Lambda_\text{UV}}
\end{align}
coupling leptons to at least one flavor of massless or very light fermions $\psi$ and a scalar $\sigma$\footnote{The presence of an ultralight scalar leads to another hierarchy problem, which we will not address here because the above is solely intended to be a toy model.}.
Here we impose a $\mathcal{Z}_2$ under which both $\psi$ and $\sigma$ are odd and that $Q_\text{X}[\psi] = 1$.  

We assume that Leptogenesis \cite{Fukugita:1986hr} is not operational in the early universe and that the baryon asymmetry is generated without involvement of the leptons, e.g. after sphaleron decoupling. Instead we consider the decay of the light active Majorana neutrinos of the cosmic neutrino background taking place in the very late universe via the operator in \eqref{eq:decay}.
Here the \enquote{Why now?} problem of dark energy is solved by kinematics, since the decay only becomes relevant when $\nu_1$ turns non-relativistic.
Note that a similar scenario for a chameleon field as early dark energy to solve the $H_0$-tension, that gets kicked into motion when neutrinos become non-relativistic, was proposed in Ref.~\cite{Sakstein:2019fmf}. 

The two heavier active neutrinos have already decayed away via $\nu_{2,3} \rightarrow \psi \sigma$, where we assume the final state particles to massless for simplicity. This could have profound implications for early universe inferences of the neutrino mass bound \cite{Escudero:2020ped,Barenboim:2020vrr}.

From the interference of the tree-level decay of the lightest active neutrino $\nu_1 \rightarrow \psi \sigma$ with the one-loop vertex- and self-energy-corrections (involving $\nu_{2,3}$ running in the loops) we can induce an asymmetry in $\psi$. The dynamics are essentially the same as for Leptogenesis, because here the massive active Majorana neutrinos play the role of the (super-)heavy decaying right handed neutrinos.  Thus $z_i$ is defined as the redshift, when the lightest massive neutrino becomes non-relativistic. 
The out of equilibrium condition for asymmetry production is given by
\begin{align}\label{eq:ooE}
  \Gamma\left(\nu_1 \rightarrow \psi \sigma,\; \overline{\psi} \sigma \right) \simeq \frac{|Y_\nu|^2 m_\nu^{(1)}}{4\pi} <H(z_i).
\end{align}
One finds that the $\psi$ asymmetry $\Delta n_\psi$ leads to an initial axion velocity of 
\begin{align}
    \dot{\theta}_i \simeq \frac{\Delta n_\psi (z_i)}{c_\psi f_a^2}
\end{align}
and the estimation of $\Delta n_\psi$ proceeds as in Leptogenesis
\begin{align}
    \Delta n_\psi = \varepsilon \kappa\; n_{\nu,1},
\end{align}
where $\varepsilon$ is the amount of $CP$-violation per active neutrino decay and $\kappa$ parameterizes the out-of-equilibrium abundance; as long as eq.~\eqref{eq:ooE} holds one obtains $\kappa \simeq 1$.  We parameterize $\varepsilon$ in terms of the  hierarchical spectrum of active neutrino masses $m_\nu$ as \cite{Davidson:2002qv} 
\begin{align}
    \varepsilon \simeq -\frac{3}{8\pi} |Y_\nu|^2 \sin(2\phi )\sum_{j=2,3} \frac{m_\nu^{(1)}}{m_\nu^{(j)}},
\end{align}
where purely for notational convenience we assumed one generation of $\psi$ and that the absolute value of the Yukawa couplings $Y_\nu$  and their irreducible phase $\phi$ are independent of the active neutrino flavor. 
Furthermore we estimate the neutrino number density of the lightest massive neutrino  in terms of the SM prediction for the amount of dark radiation $N_\text{eff.}$ that ranges between 3.043 and 3.044 \cite{Cielo:2023bqp,Drewes:2024wbw} and the photon energy density as follows $\rho_\gamma$
\begin{align}
    n_{\nu,1}(z_i) = \frac{\rho_\nu(z_i)}{3 m_\nu^{(1)}}= N_\text{eff.} \frac{7}{8} \left(\frac{4}{11}\right)^\frac{4}{3}\;\frac{ \rho_\gamma(z_i)}{3 m_\nu^{(1)}},
\end{align}
where the factor of 3 arises because we expect the  neutrino energy density $n_\nu$ to be roughly equally distributed among the  three neutrino mass eigenstates. Additionally we used the fact that for $z\geq z_i$  both energy densities redshift as radiation $\rho_\nu \sim \rho_\gamma \sim 1/R^4$ and the photon energy density is computed from its present day value of $\rho_{\gamma,0}=\SI{0.26}{\electronvolt\per\centi\meter\cubed}$ \cite{ParticleDataGroup:2022pth} as $\rho_\gamma(z_i)= \rho_{\gamma,0}(1+z_i)^4$.
The axion velocity turns out to be independent of the lightest neutrino mass and we take $z_i=10$ for concreteness
\begin{align}\label{eq:KickFerm}
    \left|\frac{\dot{\theta}_i}{m_a}\right| &\simeq \frac{7}{64 \pi} \left(\frac{4}{11}\right)^\frac{4}{3} \frac{|Y_\nu|^2 \sin(2\phi) N_\text{eff.}}{c_\psi} \sum_{j=2,3} \frac{\rho_{\gamma,0}(1+z_i)^4}{m_{\nu}^{(j)} m_a f_a^2 }\\
    &\simeq  8\times 10^{-34} \frac{|Y_\nu|^2 \sin(2\phi)}{c_\psi} \left(\frac{\SI{0.01}{\electronvolt}}{m_\nu^{(2,3)}}\right)  \left(\frac{H_0}{m_a}\right) \left(\frac{M_\text{Pl.}}{f_a}\right)^2.
\end{align}
Unfortunately  the resulting velocity will in typically  be far too small for our purposes, as the analysis in section \ref{sec:BOTH} revealed that we need $\dot{\theta}_i=\mathcal{O}(1)\; m_a$. Furthermore  small values of  $Y_\nu$ will be required to 
ensure the out of equilibrium condition in eq.~\eqref{eq:ooE}. 
This can be understood by noting that eq.~\eqref{eq:KickFerm} is essentially suppressed with respect to the velocity from a gauge field background in eq.~\eqref{eq:KickGauge} of the next section by a factor of $H_0/m_{\nu}^{(2,3)}$. This conclusion also applies to other mechanism of asymmetric dark radiation decay involving fermions, because the fermions running in the loops have to be much heavier than $H_0$. 

If one instead considers a fraction of say $m_\text{DM}^{(1)}=\mathcal{O}(\SI{100}{\electronvolt})$ fermionic DM undergoing such an asymmetric decay with  heavier $m_\text{DM}^{(2,3)}=\mathcal{O}(\SI{1}{\kilo\electronvolt})$ scale fermions running in the loop, one finds a result of slightly lower order of magnitude, since the enhancement by $\rho_m/\rho_\gamma$ is canceled by the ratio of the loop masses $m_\nu^{(2,3)}/m_\text{DM}^{(2,3)}$.

For this case one might consider a resonant enhancement of the $CP$-violating decay parameter from the self-energy diagrams by making the decaying fermion almost degenerate with the one running in the loop. We  sincerely doubt that this could bridge the required more than thirty orders of magnitude, without violating perturbativity or unitarity.

\section{Helical background of dark gauge fields}\label{sec:HelGauge}
Suppose the axion has a coupling to abelian gauge fields 
\begin{align}\label{eq:topo}
    \frac{\alpha}{8\pi} \theta F_{\mu \nu} \tilde{F}^{\mu \nu},
\end{align}
where we assume that the respective $\text{U}(1)$ is not hypercharge, but corresponds to some dark sector gauge interaction instead, and $\alpha$ is the corresponding fine structure constant. We focus on abelian gauge fields, since they do not possess instantons in flat four dimensional spacetime without dark magnetic monopoles \cite{Nomura:2015xil,Kawasaki:2017xwt,Banerjee:2024ykz}, so there is no additional contribution to the axion potential that could upset the required flatness for successful quintessence. The dark photon background survives, if there are no particles charged under the respective gauge symmetry present in the plasma (e.g. because they are heavier than the reheating temperature); otherwise it will get short-circuited by the their conductivity. One can show that the axion obtains the following velocity in this homogeneous background \cite{Kobayashi:2020bxq}
\begin{align}
    \dot{\theta} = \frac{1}{\frac{3(\omega+3)}{2}-n} \frac{\alpha}{8\pi } \frac{   F_{\mu \nu} \tilde{F}^{\mu \nu}}{H f_a^2},
\end{align}
where the scaling of $F_{\mu \nu} \tilde{F}^{\mu \nu} \sim 1/R^n$ was assumed and the scaling of the Hubble rate $H\sim 1/R^{3(\omega+1)/2}$ was utilized. We need 
\begin{align}
    n\neq 3\;(4.5)
\end{align}
for this expression to hold in a quasi de Sitter (matter dominated) background. For  massless 
gauge fields we expect the redshift behavior of radiation $n=4$.
It is evident, that  $\dot{\theta}$  has an inherent time-dependence.


The parity odd term  can be written in terms of dark electric and magnetic fields as $F_{\mu \nu} \tilde{F}^{\mu \nu} = - 4 E_\mu B^\mu$ and it is interpreted as  background of helical gauge fields, meaning that there is an asymmetry in the two circular polarizations.
The strength of the gauge field background is estimated as \cite{Kobayashi:2020bxq}
\begin{align}
|E_\mu B^\mu| \leq  \rho_A,
\end{align}
where $\rho_A =(E_\mu E^\mu + B_\mu B^\mu)/2$ is the energy density stored in the gauge field, which can be expressed in terms of its contribution to dark radiation as
\begin{align}
    \Delta N_A = \frac{8}{7} \left(\frac{4}{11}\right)^\frac{4}{3} \frac{\rho_A}{\rho_\gamma},
\end{align}
where $\rho_\gamma$ denotes the photon energy density that redshifts as $\rho_\gamma(z_i)= \rho_{\gamma,0}(1+z_i)^4$ in terms of its day value of  $\rho_{\gamma,0}=\SI{0.26}{\electronvolt\per\centi\meter\cubed}$  \cite{ParticleDataGroup:2022pth}. For $n=4$ one finds that $\Delta N_A$ is constant after electron positron annihilations  \cite{Kobayashi:2020bxq}.

For this study we choose to remain mostly agnostic about its microscopic origin, and just assume, that the helical gauge fields are produced in the late universe at e.g. $z_i=3-10$, with  fermions charged under the dark electromagnetism not being in the plasma, to ensure the survival of the gauge fields.

Using all of these ingredients and the redshift of the Hubble rate during matter domination $H(z_i)=H_0 (1+z_i)^{3/2}$ we obtain for $z_i=10$  
\begin{align}\label{eq:KickGauge}
    \left|\frac{\dot{\theta}_i}{ m_a}\right| &\leq \frac{7 }{16\pi} \left(\frac{4}{11}\right)^\frac{4}{3} \frac{\alpha \Delta N_a (1+z_i)^\frac{11}{2}}{|3-n|} \frac{\rho_{\gamma,0}}{m_a f_a^2 H_0}\\
    &\simeq 0.02 \frac{\alpha}{|3-n|} \left(\frac{\Delta N_A}{0.1}\right)   \left(\frac{H_0}{m_a}\right) \left(\frac{M_\text{Pl.}}{f_a}\right)^2 ,
\end{align}
which shows that it is hard to realize  a sufficiently large velocity of $\dot{\theta}_i=\mathcal{O}(1) \;m_a$,  even if  $\alpha$ is not too small.
Since here we assume that the dark gauge bosons are produced at late times $z_i\simeq 3$, the CMB dark radiation bound on $\Delta N_A$ does not apply and it could in principle be larger than the indicated value. 
The rolling axion will not back-react on the gauge field background as long as the following quantity stays below unity \cite{Garretson:1992vt,Anber:2006xt,Durrer:2010mq,Kobayashi:2020bxq}:
\begin{align}
    \left|\frac{\alpha}{8\pi}\frac{\dot{\theta}_i}{H}\right|  &\leq \frac{7 }{128\pi^2} \left(\frac{4}{11}\right)^\frac{4}{3} \frac{\alpha^2 \Delta N_a (1+z_i)^7}{|3-n|} \frac{\rho_{\gamma,0}}{f_a^2 H_0^2}\\
    &\simeq 0.17 \frac{\alpha^2}{|3-n|} \left(\frac{\Delta N_A}{0.1}\right)   \left(\frac{M_\text{Pl.}}{f_a}\right)^2
\end{align}
If the above is not satisfied, we expect the production of gauge fields to act as an additional source of thermal friction for the axion motion similar to the scenarios in  \cite{Berghaus:2020ekh,Berghaus:2023ypi,Berghaus:2024kra}. However it is important to point out, that the aforementioned references mainly invoke dissipation from the production of a non-abelian gauge fields known as \enquote{sphaleron-heating}, whereas we consider abelian gauge fields. Our quintessence field, that gets pushed up its potential by a helical gauge field background, is in a sense complementary to a quintessence field with friction from the production of gauge fields \cite{Berghaus:2020ekh,Berghaus:2023ypi,Berghaus:2024kra}.

So far we have not specified the origin of the dark electromagnetic background.   
A possible scenario could be the decay of asymmetric fermionic dark matter to a dark photon and a chiral fermion (similar to the radiative decay of a sterile neutrino to a photon plus an active neutrino), which was argued in Ref.~\cite{Yin:2024trc} to produce polarized vector bosons. Since the dark matter and the chiral fermion will only couple to dark photons via loops, their couplings can be suppressed enough to avoid a large conductivity of the plasma. 

It is important to ensure that the axion does not decay to massless dark photons via the interaction in \eqref{eq:topo}, but this is easily avoided due to the large  separation of scales between $m_a$ and $f_a$, which can be seen from the following estimate for the axion lifetime in
\begin{align}
    \frac{\tau_a}{\tau_\text{universe}}\simeq \frac{10^{122}}{\alpha^2} \left(\frac{H_0}{m_a}\right)^3 \left(\frac{f_a}{M_\text{Pl.}}\right)^2
\end{align}
in units of the age of the universe $\tau_\text{universe}\simeq 1/H_0$.
We do not expect  nonperturbative effects due to Bose enhancement \cite{Alonso-Alvarez:2019ssa} to change this conclusion.

\bibliographystyle{JHEP}
\bibliography{refsDE}
\end{document}